%% file: vmc_fpga.tex
\pdfoutput=1
\documentclass[aps,amssymb,twocolumn,superscriptaddress]{revtex4-1}

\usepackage{amsfonts,amsmath,amsbsy,color}
\usepackage[]{graphics,graphicx}
\usepackage[version=3]{mhchem}
\usepackage{listings}
\usepackage{algorithm}
\usepackage[noend]{algpseudocode}
\usepackage{tikz,pgfplots}
\usepackage{subcaption}

\usetikzlibrary{arrows,positioning,calc,intersections,backgrounds}
\usetikzlibrary{shadows.blur}
\usetikzlibrary{shapes.symbols}
\usetikzlibrary{shapes.multipart,shapes.geometric}

\usepgfplotslibrary{external}

\definecolor{myred}{RGB}{242,21,21}
\definecolor{myblue}{RGB}{21,21,161}
\definecolor{mygreen}{RGB}{21,161,21}
\definecolor{myyellow}{RGB}{218,165,32}
\newcommand*\chem[1]{\ensuremath{\mathrm{#1}}}
\newcommand\dx[1]{\mathrm{d}{#1}}
\renewcommand{\vec}[1]{\boldsymbol{#1}}       

\begin{document}

\title[]{Field-Programmable Gate Arrays and Quantum Monte Carlo:\\ Power Efficient Co-processing for Scalable High-Performance Computing}
\author{Salvatore Cardamone}
\email{sc2018@cam.ac.uk}
\affiliation{University Chemical Laboratory, Lensfield Road, Cambridge CB2 1EW, United Kingdom}
\author{Jonathan~R.~Kimmitt}
\affiliation{University Chemical Laboratory, Lensfield Road, Cambridge CB2 1EW, United Kingdom}
\affiliation{Department of Computer Science and Technology, William Gates Building, J. J. Thomson Ave, Cambridge CB3 0FD, United Kingdom}
\author{Hugh~G.~A.~Burton}
\affiliation{University Chemical Laboratory, Lensfield Road, Cambridge CB2 1EW, United Kingdom}
\author{Alex~J.~W.~Thom}
\affiliation{University Chemical Laboratory, Lensfield Road, Cambridge CB2 1EW, United Kingdom}
\email{ajwt3@cam.ac.uk}

\date{\today}

\begin{abstract}
	Massively parallel architectures offer the potential to significantly accelerate an application relative to their serial counterparts. However, not all applications exhibit an adequate level of data and/or task parallelism to exploit such platforms. Furthermore, the power consumption associated with these forms of computation renders ``scaling out'' for exascale levels of performance incompatible with modern sustainable energy policies. In this work, we investigate the potential for field-programmable gate arrays (FPGAs) to feature in future exascale platforms, and their capacity to improve performance per unit power measurements for the purposes of scientific computing. We have focussed our efforts on Variational Monte Carlo, and report on the benefits of co-processing with an FPGA relative to a purely multicore system.
\end{abstract}

\maketitle

\section{Introduction}
Quantum Monte Carlo (QMC) encompasses a class of techniques for approximating expectation values to quantum mechanical observables for many-electron systems\cite{foulkes2001quantum}. By casting the time-independent Schr\"{o}dinger equation into integral form, the high-dimensional integration --- intractable through quadrature techniques --- can be realised through a stochastic sampling of the many-electron wavefunction.\\
Relative to deterministic counterparts, QMC offers the potential to combine the favourable scaling of mean-field techniques and the accuracy of post-Hartree Fock methods\cite{needs2009continuum}. Two forms of QMC in particular have gained traction as popular \textit{ab initio} methodologies\cite{hammond1994monte}: Variational Monte Carlo (VMC) and Diffusion Monte Carlo (DMC). Commonly, VMC is used to variationally optimise free parameters in the wavefunction, with DMC being subsequently applied to compute production level observables for the molecular system in question.

Prior to the failure of Dennard scaling\cite{dennard1974design}, application developers were able to indulge in a ``free lunch'', whereby they could exploit the continual increase in processor clock frequencies with new generations of hardware to directly accelerate an application. Since on-chip power densities no longer scale down with transistor density, manufacturers instead look towards heat dissipation technologies in an effort to accommodate Moore's law\cite{kaeli2015heterogeneous}. However, such techniques have limitations and eventually require manufacturers to reduce on-chip voltages, forecasted to eventually culminate in an era of ``dark silicon''\cite{esmaeilzadeh2011dark}.\\
To address the unsustainability of frequency scaling, the tendency has been for new generations of hardware to increase the number of parallel compute units on a single chip. As a result, the onus falls onto the software developer to exploit either operations or tasks that can be undertaken in parallel within an application. Furthermore, with the advent of distributed memory systems and graphical processing units (GPUs) at contemporary high-performance computing (HPC) facilities, the software developer has an enormous arsenal of hardware solutions for exploiting parallelism within an application.\\

For QMC methodologies, the associated embarrassing parallelism lends itself well to modern parallel architectures. There is a significant body of work demonstrating the near-linear scaling of performance with respect to the available hardware concurrency\cite{shulenburger2011hybrid,kim2018qmcpack}. More recently, GPUs have been utilised as co-processors to deliver impressive performance gains\cite{esler2012accelerating}. However, ``scaling out'' is not a sustainable solution to simulating increasingly complex systems. The power consumption associated with distributed memory systems and GPUs does not align with modern energy policies. Indeed, a consensus appears to have emerged\cite{amarasinghe2011ascr,shalf2010exascale,bergman2008exascale} that the power consumption of HPC is one of the most prodigious barriers to attaining exascale levels of performance.\\
Field-Programmable Gate Arrays (FPGAs) are appealing candidates as processing units for novel exascale platforms\cite{stroobandt2016extra,ciobanu2015extra}. These devices combine low power consumption with the capacity to exploit both data and task parallelism, and are therefore applicable to a wider set of applications than solutions relying on data parallelism alone. To date, use of FPGAs in electronic structure theory has been somewhat limited, although several works demonstrate the power of this platform to accelerate scientific applications within this domain.\cite{gothandaraman2008fpga,cooper2017quantum}\\

In this work, we look to port the compute-intensive portions of a VMC calculation to an FPGA and assess the performance relative to a CPU-bound application. We elaborate on the optimisations incorporated into our design to optimise the implementation for the purposes of co-processing. Finally, we demonstrate that our application benefits from the use of FPGAs in terms of both raw compute performance and power consumption.

\section{Variational Monte Carlo}

The following section is not intended to present an exhaustive exposition of techniques within QMC. Rather, the reader is directed to the excellent review of Foulkes and coworkers\cite{foulkes2001quantum} for further details.

\subsection{Implementation}

Consider a molecular system comprising $N$ nuclei and $n$ electrons, where $\vec{R}$ and $\vec{r}$ denote their collective position vectors respectively. The time-independent electronic wavefunction of this system is represented by $\Psi(\vec{r};\vec{R})$, where the parametric dependence on $\vec{R}$ arises from the application of the Born--Oppenheimer approximation. Expanding $\Psi(\vec{r};\vec{R})$ in terms of the exact (orthonormalised) eigenfunctions of the molecular Hamiltonian, $\hat{\mathcal{H}}(\vec{r},\vec{R})$, the energy of the system is determined by the expectation value $\langle \hat{\mathcal{H}}(\vec{r},\vec{R}) \rangle_{\Psi}$. Alternative (time-independent) physical quantities can be obtained by substituting the molecular Hamiltonian for the appropriate corresponding operator.\\
From the variational principle it can be shown that for any approximate wavefunction, $\Psi_T(\vec{r};\vec{R})$, referred to as the ``trial wavefunction'', the expectation value $\langle\hat{\mathcal{H}}(\vec{r},\vec{R})\rangle_{\Psi_T}$, provides an upper bound to the true energy of the system. Consequently, the variational principle provides an important metric for quantifying the quality of an approximate trial wavefunction, with lower energies implying a higher quality wavefunction. Herein, all explicit functional dependencies are omitted unless a new quantity is introduced.\\
Through the time-independent Schr\"{o}dinger equation, the energy associated with a particular trial wavefunction can be expressed as
\begin{equation}
	E_T = \frac{ 
			\int \dx{\vec{r}} \Psi_T^*\hat{\mathcal{H}}\Psi_T
		}{
			\int \dx{\vec{r}} \Psi_T^*\Psi_T
		}\,,
\end{equation}
where integration is performed over the $3n$ electronic degrees of freedom. With some foresight, multiplication by the identity $\Psi_T/\Psi_T$ casts this expression into a form amenable to solution through stochastic methods:
\begin{equation}
	E_T = \frac{ 
		\int \dx{\vec{r}} |\Psi_T|^2 \frac{\hat{\mathcal{H}}\Psi_T}{\Psi_T}
	}{
		\int \dx{\vec{r}} |\Psi_T|^2
	} \,.
\end{equation}
The above ``importance sampled form'' utilises $|\Psi_T|^2$ as a probability density function from which samples of the ``local energy'', $E_L(\vec{r}) = {\Psi_T}^{-1} \hat{\mathcal{H}}\Psi_T$, can be drawn. Through Monte Carlo, the resultant average of the local energy over $N_{\textrm{MC}}$ (uncorrelated) samples is asymptotic to the variational energy of the trial wavefunction:
\begin{equation}
	E_T \sim \lim_{N_{\textrm{MC}}\rightarrow\infty} \frac{1}{N_{\textrm{MC}}} \sum_{i=1}^{N_{\textrm{MC}}} E_L(\vec{r}^{(i)}) \,,
\end{equation}
where $r^{(i)}$ denotes a sample, and equality emerges in the limit $N_{\textrm{MC}}\rightarrow\infty$.\\
Variational Monte Carlo (VMC) provides a means for implementing the above stochastic process. Considering an ensemble of ``walkers'', each representing a discrete sample of $\Psi_{\mathrm{T}}$ with a particular electronic configuration in position space, the simulation proceeds by stochastically propagating each walker through configuration space.
By randomly displacing an electronic configuration, $\vec{r}^\prime \leftarrow \vec{r}$, the Metropolis--Hastings algorithm can be invoked to accept the step according to the transition probability
\begin{equation}
	P(\vec{r}^\prime\leftarrow\vec{r}) = \min\left[ 1, \Bigg\lvert \frac{ \Psi_T(\vec{r}^\prime)}{ \Psi_T(\vec{r}) } \Bigg\rvert^2 \right] \,.
\end{equation}
A basic overview of the VMC algorithm follows in Algorithm \ref{alg::VMC}.

\begin{algorithm}[H]
\begin{algorithmic}

	\For{ $\texttt{iMC} = 1,\hdots,N_{\textrm{MC}}$ }
		\For{ $\texttt{iWalker} = 1,\hdots,N_w$ }
			\For{ $\texttt{iEl} = 1,\hdots,n$ }

				\State $\vec{r}^\prime_{\texttt{iEl}} \gets \vec{r}_{\texttt{iEl}}+\vec{\delta}$
				\State Compute $\Psi_T(\vec{r}^\prime)$
				\If{ $|\Psi_T(\vec{r}^\prime) / \Psi_T(\vec{r})|^2 > \mathcal{U}(0,1)$ }
					\State Compute $\nabla^2_{\vec{r}^{\prime}}\Psi_T(\vec{r}^\prime)$
					\State Update $\Psi_T^{-1}(\vec{r}^\prime)$
					\State Accumulate $E_L(\vec{r}^\prime)$
    					\State $\vec{r}_{\texttt{iEl}} \leftarrow \vec{r}^\prime_{\texttt{iEl}}$				
				\EndIf
			\EndFor
		\EndFor
	\EndFor

	\Return $E_T$
	
\end{algorithmic}
\caption{Variational Monte Carlo}
\label{alg::VMC}
\end{algorithm}

\subsection{The Trial Wavefunction}

Since the many-electron wavefunction is unknown in closed-form for all but the most trivial of systems, approximations must be invoked. Owing to the antisymmetry of the fermionic wavefunction, a determinant is a particularly appropriate mathematical form for the trial wavefunction. Subject to the spin-invariance of the operator whose associated observable is to be computed, the trial wavefunction can be written as the product of $\alpha$- and $\beta$-spin components,
\begin{equation}
	\Psi_T(\vec{r}^\alpha,\vec{r}^\beta) = \det[ \mathcal{D}^\alpha(\vec{r}^\alpha) ] \times \det[ \mathcal{D}^\beta(\vec{r}^\beta) ] \,,
\end{equation}
where $\mathcal{D}(\vec{r})$ is referred to as the Slater matrix, and $\vec{r}^\alpha,\vec{r}^\beta$ denote the set electron position with the appropriate spin. The Slater matrix is composed of a set of one-particle molecular orbitals, $\psi(\vec{r})$,
\begin{equation}
	\mathcal{D}^\omega(\vec{r}^\omega) = 
	\left[\begin{array}{ccc}
		\psi_1(\vec{r}_1^\omega) & \hdots & \psi_{n_\omega}(\vec{r}_1^\omega) \\
		\vdots & \ddots & \vdots \\
		\psi_1(\vec{r}_n^\omega) & \hdots & \psi_{n_\omega}(\vec{r}_n^\omega)
	\end{array}\right] \,,
\end{equation}
where $\omega$ symbolises an arbitrary spin-state. A molecular orbital is constructed as a linear combination of atomic orbitals (LCAO),
\begin{equation}\label{eq::LCAO}
	\psi_i(\vec{r}) = \sum_{j=1}^{N_{\textrm{AO}}} c_{ij}\phi_j(\vec{r}) \,,
\end{equation}
where $\phi(\vec{r})$ denotes an atomic orbital, $N_{\textrm{AO}}$ is the number of atomic orbitals in the expansion and $\{c_{ij}\}$ are the expansion coefficients for the $i^{\textrm{th}}$ molecular orbital. The atomic orbital is formed from a linear combination of $N_p$ primitive functions,
\begin{align}\label{eq::Primitives}
	\phi_j(\vec{r}) &= f(\ell_j,m_j,|\vec{r} - \vec{R}_j|) \sum_{k=1}^{N_p} d_{jk} \exp( -\zeta_k|\vec{r} - \vec{R}_j|^2) \nonumber \\
	&= f(\ell_j,m_j,\vec{r}) \sum_{k=1}^{N_p} \eta_{jk}(\vec{r}) \,.
\end{align}
In the above, $\vec{R}_j$ is the position vector of the atom to which the atomic orbital is centred, $\zeta_k$ is the width of the gaussian function and $d_{jk}$ are the expansion coefficients for the $j^{\textrm{th}}$ atomic orbital. $f(\ell_j,m_j,\vec{r})$ is a function of the azimuthal and colatitudinal quantum numbers associated with the atomic orbital, i.e. its angular momentum.\\

It is common practice to multiply the above determinantal form of the wavefunction with a function of inter-particle distances, referred to as the Jastrow factor, to account for the correlation effects between particles\cite{drummond2004jastrow}. While inclusion of the Jastrow factor is one of the powerful features of QMC methods in general, we omit discussion of it in our work for the sake of simplicity. Furthermore, the sum of gaussian functions arising in the expression for the molecular orbital is often substituted for a cubic spline to ease the computational burden of complexity scaling with the number of atomic orbitals in the system\cite{williamson2001linear}. We avoid use of these splines to ensure our application remains compute bound.\\
We wish to stress that our resultant implementation is optimised for a subset of VMC capabilities. Rather than claim our results are representative of all VMC calculations, we hope this work serves to illustrate the potential benefits of using FPGAs, and as such should be considered exploratory as opposed to all-encompassing.
 
\section{FPGAs}

\subsection{Basics and Nomenclature}

\begin{figure*}[t]
	\begin{subfigure}[t]{.48\textwidth}
		\tikzsetnextfilename{unpipelined}
		\scalebox{.65}{\input{unpipelined2.tikz}}
		\caption{\raggedright Unpipelined three-module workflow, each module being dependent upon the output of the former in sequence. A single data element must traverse the entire workflow prior to the entry of the next data element. While the latency is comparatively lower than for a pipelined case, the throughput is also low.}.
		\label{fig::PipeliningA}
	\end{subfigure}
	\begin{subfigure}[t]{.48\textwidth}
		\tikzsetnextfilename{pipelined}
		\scalebox{.65}{\input{pipelined.tikz}}
		\caption{\raggedright Pipelined three-module workflow, each module being dependent upon the output of the former in sequence. Multiple data elements are allowed to reside within the workflow simultaneously. The latency of the workflow is higher than the unpipelined case owing to the necessity that all modules have associated registers to increase the \textit{effective} module latency to that of the highest module latency. However, throughput is twice as high as for the unpipelined case, with a new result output every $\ell_g$ once the pipeline is filled.}
		\label{fig::PipeliningB}
	\end{subfigure}
	\caption{\raggedright An application comprising a series of data-dependent modules and its amenability to optimisation through pipeline parallelism.}
	\label{fig::Pipelining}
\end{figure*}
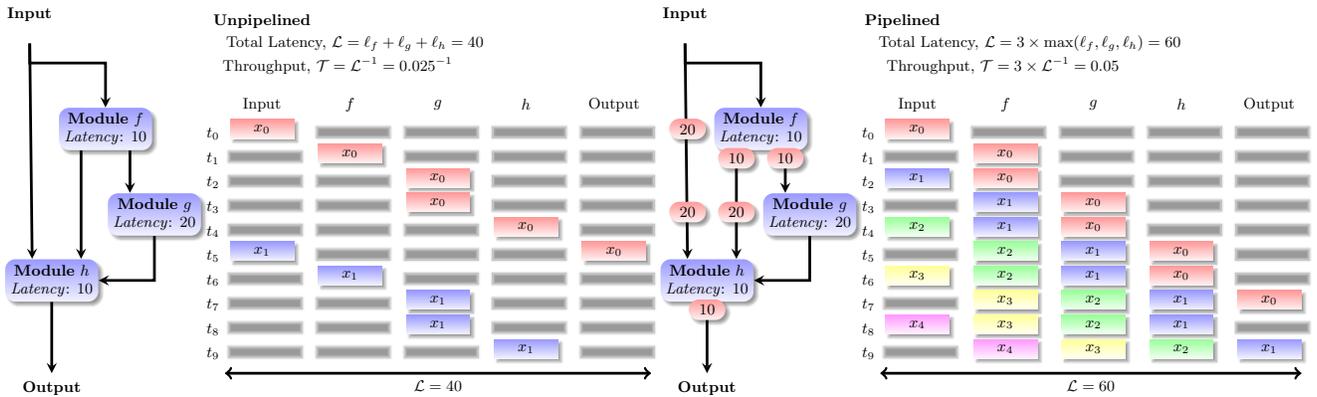

From the beginning it is worth clarifying the distinction between \textbf{latency} and \textbf{throughput}, the former being the time taken to traverse a computational workflow, and the latter being the number of outputs from the workflow per unit time; for the processing of a single data item, these are just reciprocally related. For multiple data items, pipeline parallelism offers the capacity to increase computational throughput at the cost of latency. The rationale behind leveraging latency for throughput is best illustrated by example, for which we will refer to Figure \ref{fig::Pipelining}.\\
Consider a ``stream'' of data, $x_0,x_1,\hdots$, where the subscript enumerates order. In Figure \ref{fig::PipeliningA}, we observe an unpipelined implementation of three data-dependent modules: $f$, $g$ and $h$, each delivering a result through composition with the preceding modules, i.e. $f(x)$, $g(x,f)$ and $h(x,f,g)$, where nested parentheses are omitted for clarity. Each module has an associated latency, $\ell_f, \ell_g, \ell_h$. The latency of this workflow is given by the longest route through the workflow, so here is simply the sum of the individual module latencies. The maximum frequency at which a clock can drive the workflow is determined by the propagation delay associated with the longest pathway in the workflow. As the module latencies differ, allowing multiple elements of the data stream to concurrently reside within the workflow results in improper behaviour. The occupancy of data within modules of the workflow with respect to time is depicted in the right-hand side of Figure \ref{fig::PipeliningA}.\\
Figure \ref{fig::PipeliningB} depicts the same workflow as that in \ref{fig::PipeliningA}, but with the addition of registers (red boxes and their associated latencies) capable of storing data, across the workflow. As such, each module effectively inherits its latency from the maximum module latency in the workflow, $\ell_g$ in this case. However, introduction of the registers allows the workflow to process multiple data concurrently. In other words, the data stream exhibits temporal, or \textbf{pipeline}, parallelism, i.e. the task can be executed as a cascade of sub-tasks\cite{freitas2000basic}. The resultant pipelined implementation then increases the throughput of the application, at the cost of a delay in processing a single datum.\\

An FPGA grants the application developer the means to realise a computational pipeline spatially in silicon. The FPGA is a matrix of configurable logic blocks (CLBs), fundamental programmable units comprising a lookup table and flip-flop (a logic unit capable of storing a state), amongst other logical units depending upon the chip. Signals are routed through the CLBs by a series of programmable switches, theoretically permitting the transmission of a signal between any two CLBs (although in practice one would wish to co-localise the connected CLBs for an optimal configuration). Typically, a number of ``hard blocks'' (such as digital signal processors, dedicated floating point units, static random access memory (SRAM) blocks, etc.) are also available on the chip for specialised tasks that may be costly to implement directly from CLBs.\\
A configuration for the FPGA is loaded at runtime into a flash memory. The contents of this memory are used to configure the programmable switches, routing signals as per the uploaded configuration. For the types of computation considered in this work, the FPGA is able to interact with a general purpose processing unit through, for instance, a PCIe connection. Since the FPGA is configured to perform a specific set of computations, the overheads associated with general purpose computing (such as scheduling and interrupts controlled by an operating system) are eliminated. Furthermore, since the clock frequency at which an FPGA configuration can be run is dictated by the propagation delay across the chip, the implementation will be clocked at far lower frequency than a CPU or GPU. These two factors result in a dramatic decrease in the power consumption of an FPGA. As a rough indication, an FPGA consumes roughly an order of magnitude less power than a CPU or GPU. As such, FPGAs are attractive devices for the purposes of low-power computation.\\

Naturally there are some fairly sizeable obstacles to the use of FPGAs. The conventional means for programming FPGAs requires knowledge of low-level unabstracted hardware description languages (HDLs). Use of these languages requires expertise in low-level design, and is therefore typically inaccessible to application developers from the physical sciences. Consequently, development times for FPGA-based applications are enormously lengthy.\\
However, there exist a number of tools available to the developer for porting a complex application to FPGAs. High Level Synthesis (HLS) tools, such as those marketed by the major FPGA vendors Xilinx and Intel/Altera, offer the capacity to annotate high-level source code (C/C++ and OpenCL predominantly) with preprocessor directives, subsequently used to construct a HDL implementation of the application. Nevertheless, high performance implementations require considerable restructuring relative to a conventional CPU-based alternative to facilitate the inference of a pipelined implementation by the HLS. While the level of expertise required by the developer is reduced, a significant understanding of the platform is imperative.\\
An alternative approach is to use an embedded domain specific language (EDSL), providing a library for a high-level host language to support FPGA-based constructs, such as streams of data. The implementation is analysed to construct an abstract syntax tree, which can subsequently be used to generate the HDL implementation of the application. An example of such a facility is the extension to java provided by Maxeler Technologies, of which more will be said later in this work. Similar to HLS, an EDSL requires that the developer write their application in a form amenable to constructing a FPGA implementation, and consequently still requires a knowledge of the platform.

\subsection{Application Overview}

Gothandaraman and coworkers\cite{gothandaraman2008fpga} have previously ported a VMC application to an FPGA, directing their efforts to potential energy and trial wavefunction evaluation kernels. Through the use of pipelining and fixed point arithmetic, significant improvements in performance were obtained relative to a serial multicore implementation. However, their work was limited to bosonic systems, the trial wavefunction consequently being expressible as a product of functions of pairwise particulate distances, i.e. there is no need to invoke a determinantal form for the trial wavefunction. We have targeted fermionic VMC, and as a result our implementation requires significant departures from previous efforts.\\ 

To fully exploit the reconfigurability of an FPGA, we have chosen to write an in-house VMC code so as to not restrict ourselves to the data structures and algorithmic workflows utilised in more sophisticated packages\cite{needs2009continuum,wagner2009qwalk,kim2018qmcpack}. Rather, we have been able to write the application so as to optimise computation through co-processing with an FPGA. It is worth reiterating that we consider here only a subset of VMC functionality; our implementation is all-electron and Jastrow-free.\\
\begin{table*}
\caption{\raggedright Wall time and percentage of cache misses (from the number of cache references) for the contiguous and triplet data structures (first two columns), and parallelising the $\texttt{iAt}$ and $\texttt{jAt}$ loops of the pairwise distance kernel using OpenMP multithreading. Data obtained from unique pairwise distances between 100,000 particles. Multithreading data obtained using two threads with the contiguous data structure.}
\label{tbl::Cache}
\begin{ruledtabular}
\begin{tabular}{c|l|c|c|c}
	\centering
	& \textbf{Contiguous} & \textbf{Triplet} & \textbf{$\texttt{iAt}$ OpenMP} & \textbf{$\texttt{jAt}$ OpenMP} \\
	\hline
	Wall Time (s) & 20.107 & 22.831 & 15.936 & 13.057 \\
	Cache Misses & 0.279\% & 0.534\% & 16.479\% & 0.594\% \\
\end{tabular}
\end{ruledtabular}
\end{table*} 
We have aspired to optimise our CPU implementation so as to possess a reasonable benchmark. Our VMC code is written in ISO C99, with all data structures written in a ``struct of arrays'' (SoA) format. The application is multithreaded using OpenMP. We have attempted to replicate the benefits of code encapsulation offered by object-oriented approaches; a struct is utilised as a means for storing data and function pointers. The $\texttt{Ensemble\_t}$ struct, for instance, comprises electron positions, Slater matrices, laplacians, and all other data one may associate with a walker, along with methods for manipulating these data.\\ 
Our application foregoes physically motivated data structures to facilitate optimal cache behaviour\cite{mathuriya2017embracing}. Such optimisations require a significant effort on the part of the developer,  but their implementation can help to ameliorate an application from becoming overly memory-bound. Consider, for instance, a set of $n$ particles, each described by a position vector in $\mathbb{R}^3$. One can conceive of two separate data formats: the contiguous dimensions, i.e. $\{x_1,\hdots,x_n,y_1,\hdots,y_n,z_1,\hdots,z_n\}$; and the physically motivated triplets, $\{x_1,y_1,z_1,x_2,y_2,z_2,\hdots,x_n,y_n,z_n\}$. An example of explicit computation of the unique pairwise square distances between particles is given in Algorithm \ref{alg::PairwiseTriplet}. 

\begin{algorithm}[H]
\scriptsize{
\begin{lstlisting}[language=C++]
static const size_t nAtoms = 100000, nDims = 3 ;
static float positions[nAtoms*nDims] ;	
static size_t xi, yi, zi = 0 ;	
static size_t xj, yj, zj = 0 ;	

for( size_t iAt=0 ; iAt<nAtoms ; ++iAt ) {

#ifdef CONTIGUOUS
   xi = iAt ; yi = nAtoms + xi ; zi = nAtoms + yi ;
#else /* TRIPLET */
   xi = iAt * nDims ; yi = xi + 1 ; zi = yi + 1 ;	
#endif /* #ifdef CONTIGUOUS */
	
   for( size_t jAt=iAt+1 ; jAt<nAtoms; ++jAt ) {

#ifdef CONTIGUOUS	
      xj = jAt ; yj = nAtoms + xj ; zj = nAtoms + yj;	
#else /* TRIPLET */
      xj = jAt * nDims ; yj = xj + 1 ; zj = yj + 1 ;	
#endif /* #ifdef CONTIGUOUS */	

      float dx = positions[xi] - positions[xj] ;	
      float dy = positions[yi] - positions[yj] ;	
      float dz = positions[zi] - positions[zj] ;	

      float rSq = dx*dx + dy*dy + dz*dz ;		
		
   }
}	
\end{lstlisting}
}
\caption{Pairwise Distances using Contiguous or Triplet Data Structures}
\label{alg::PairwiseTriplet}
\end{algorithm}
The first two columns of Table \ref{tbl::Cache} give the number of cache loads and associated proportion of cache misses from use of the two data formats. The contiguous case gives significantly improved cache performance since the square distance between particles is computed dimension-at-a-time, i.e. one computes $\texttt{dx,dy,dz}$ in sequence, prior to reduction and computation of the square root. Allowing the compiler to vectorise the code, one finds that the implementation effectively computes $\texttt{dx}$ between $\texttt{iAt}$ and several $\texttt{jAt}$ concurrently. Since the $x-$degrees of freedom are contiguous, a single fetch is adequate to serve the vectorised operation. In contrast, for the triplet storage, one finds that a number of elements in the cache are redundant for the calculation of $\texttt{dx}$, i.e. the $y,z$ degrees of freedom. As such, the cache does not contain all of the required data to serve the vectorised instruction, resulting in suboptimal caching.\\
The final two columns then consider the multithreaded implementation of the pairwise distance kernel given the contiguous data structure. The cases differ based on which loop is parallelised: $\texttt{iAt}$ or $\texttt{jAt}$. Given parallelisation of the $\texttt{iAt}$ loop, one finds each thread attempting to load its assigned $\texttt{iAt}$ simultaneously, along with corresponding $\texttt{jAt}$. The result can lead to exceedingly poor performance as a result of cache-thrashing. On the other hand, parallelising the $\texttt{jAt}$ loop results in each thread working on a single $\texttt{iAt}$, and the cache contains all required data for the parallelisation over $\texttt{jAt}$.\\

\begin{figure*}[]
	\tikzsetnextfilename{host_device}
	\scalebox{1.5}{\input{host_device.tikz}}
	\caption{\raggedright Schematic of the host-device architecture. A host is a regular multicore CPU, all processing units having access to some bank of main memory. The device is able to interact with the host through PCIe. Memory transactions between host and device can take place between the main memory of the host and the FPGA, which can either utilise the data directly, or route it to device DRAM.}
	\label{fig::HostDevice}
\end{figure*}
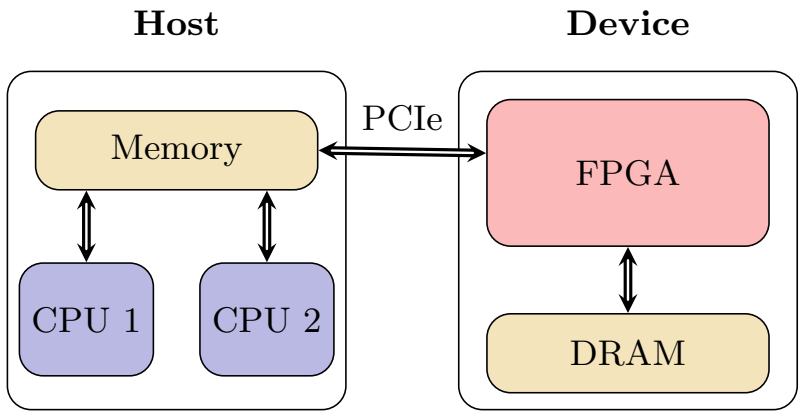

In the following section, we adopt the conventional nomenclature ``host'' and ``device'' to denote the CPU and the FPGA, respectively (see Figure \ref{fig::HostDevice}). The device is connected to the host through PCI express (PCIe), permitting memory transfers to proceed by direct memory access (DMA), i.e. without blocking the CPU. Owing to the fact that VMC is embarrassingly parallel, it should hold that the difference in performance between a host and a host with device is conserved in a multiprocessing environment, with $n$ hosts and $n$ devices. Our study then compares single host against single host with device.\\
Our FPGA implementation is written in the EDSL provided by Maxeler Technologies, maxj. maxj is an extension to java, providing functionalities for dealing with sequences of data which are streamed across an FPGA implementation of the application. Such an implementation is referred to as ``dataflow''. maxj is interpreted by the MaxCompiler, which constructs a register transfer level (RTL) implementation of the application. The RTL is subsequently passed to a synthesis tool (supplied by the chip vendor) to construct the FPGA configuration.\\ 

\section{Implementation Details}

Throughout, we use the MAIA board produced by Maxeler Technologies, possessing an Altera Stratix V FPGA chip and 48GB of on-board DDR3 DRAM. Our host processor is an Intel Xeon E5-2640 with 6 physical cores, clocked at 2.5GHz. The molecular system we work with is a lattice of 64 molecules of $\chem{H}_2$. We choose the STO-6G basis set for the sake of simplicity, although our described implementation is general enough to utilise an arbitrary basis set.\\
Evaluating the trial wavefunction and its derivatives is the major computational bottleneck of VMC. For small systems such as $\chem{Be_2}$, profiling of our code reveals over 70\% of the compute time is localised in the routines which evaluate atomic and molecular orbitals. As the system size increases, the trial wavefunction routines dominate further: upwards of 90\% of compute time for the lattice of hydrogen. As such, our target for offloading to the FPGA is obvious.\\
While the Sherman--Morrison inverse (and determinant) updating routine accounts for a non-negligible portion of runtime, it has been noted that the routine is fairly efficient on CPUs\cite{esler2012accelerating} since the inverse matrices fit comfortably in cache. We also note that an improved inverse updating scheme has recently been published\cite{mcdaniel2017delayed}. In the future, it may be interesting to incorporate these routines into an FPGA implementation. However, for the purposes of this work, we concentrate our efforts specifically on the trial wavefunction evaluation kernel.\\

\subsection{Workflow Outline}

The algorithmic workflow outlined in Algorithm \ref{alg::VMC} serves to illustrate the data dependencies of the individual kernels, and the kernels which must be serialised. An electron must be displaced, and the trial wavefunction subject to this displacement evaluated. Should the new trial wavefunction meet the Metropolis criterion, the inverse and laplacian of the trial wavefunction are computed, and the local energy accumulated. Should the trial wavefunction not meet the Metropolis criterion, the electron is reverted to its position prior to displacement.\\
There is no scope for task-level parallelism, rendering it difficult to occupy both host and device simultaneously. Simply offloading the computationally intensive portions of the calculation to the device and blocking the CPU until completion is inefficient from a co-processing perspective. The offload cost must be taken into account when benchmarking against a host-bound application. Any savings made by the co-processor in calculation time must be adjusted for the overhead associated with the offload.\\
The embarrassingly data-parallel nature of our application can be used to mask the offload cost. By simply splitting our ensemble into two equally-sized ensembles, \texttt{Ensemble\_A} and \texttt{Ensemble\_B}, one can be processed by the host while the other is processed by the device. In effect, we artificially create pipeline-parallelism through data-parallelism. However, this enforced task-parallelism comes at the price of additional memory consumption.\\
Consider momentarily \texttt{Ensemble\_A} in isolation of \texttt{Ensemble\_B}. At $t_0$, it is offloaded to the DFE, where the constituent electrons are displaced, the Slater matrix differences and laplacians are computed, and subsequently sent back to the host. The host is now holds \texttt{Ensemble\_A} at $t_1$, where all trial moves have speculatively been accepted. Upon proceeding to the Metropolis step on the host, one finds that upon rejection of a particular move, the data associated with the previous configuration has been overwritten, resulting in the requirement that the host recompute the data from $t_0$.\\
Now, consider the case where the data associated with \texttt{Ensemble\_A} is double-buffered. The data at $t_0$ is stored in both the ensemble's buffers. The double-buffer is offloaded to the device which again speculatively updates \texttt{Ensemble\_A} to $t_1$, the data for which is subsequently passed back to the double-buffer. Now, when attempting the Metropolis step, the host has access to data from both $t_0$ and $t_1$, allowing either rejection or acceptance of a trial move without the need to recompute data.\\
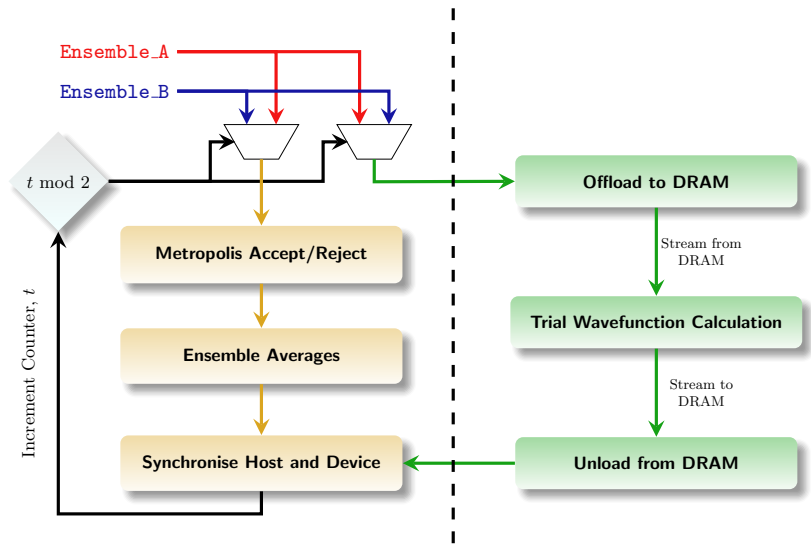
\begin{figure*}[t]
	\tikzsetnextfilename{non_blocking}
	\scalebox{1.5}{\input{non_blocking.tikz}}
	\caption{\raggedright Non-Blocking implementation of VMC. The dashed line denotes the partition between host (yellow) and device (green) kernels. The destination of the two ensembles alternates between counter ticks, and can be handled by simple ternary logic (represented by multiplexing).}
	\label{fig::NonBlockingWorkflow}
\end{figure*}
Finally, let us reintroduce \texttt{Ensemble\_B}. Now, while \texttt{Ensemble\_A} is being processed by the device, \texttt{Ensemble\_B} can undergo the Metropolis step on the host having previously been speculatively updated on the device. The two ensembles can subsequently alternate being processed by the host and device, resulting in a non-blocking simulation. Furthermore, the host is free to run multithreaded, and the device need never block the host since memory transfers can proceed through DMA (unless the host is undertaking some memory-intensive task, in which case the southbridge will be overfaced). The number of threads the host uses can be tuned to minimise the amount of idle time and power consumed. An overview of this workflow is given in Figure \ref{fig::NonBlockingWorkflow}.

\subsection{Dataflow Trial Wavefunction Evaluation}

Owing to the enormous computational efforts associated with hardware synthesis, our main aim the construction of a generic dataflow implementation for the trial wavefunction kernel. By generic, we mean that the implementation should be able to process a system of arbitrary complexity (within reason), without the need for resynthesis. Naturally, some constraints must be placed on the FPGA configuration such that it does not simply exhaust all resources on-chip. However, we attempt to minimise these constraints in an attempt to maintain freedom in the design space.
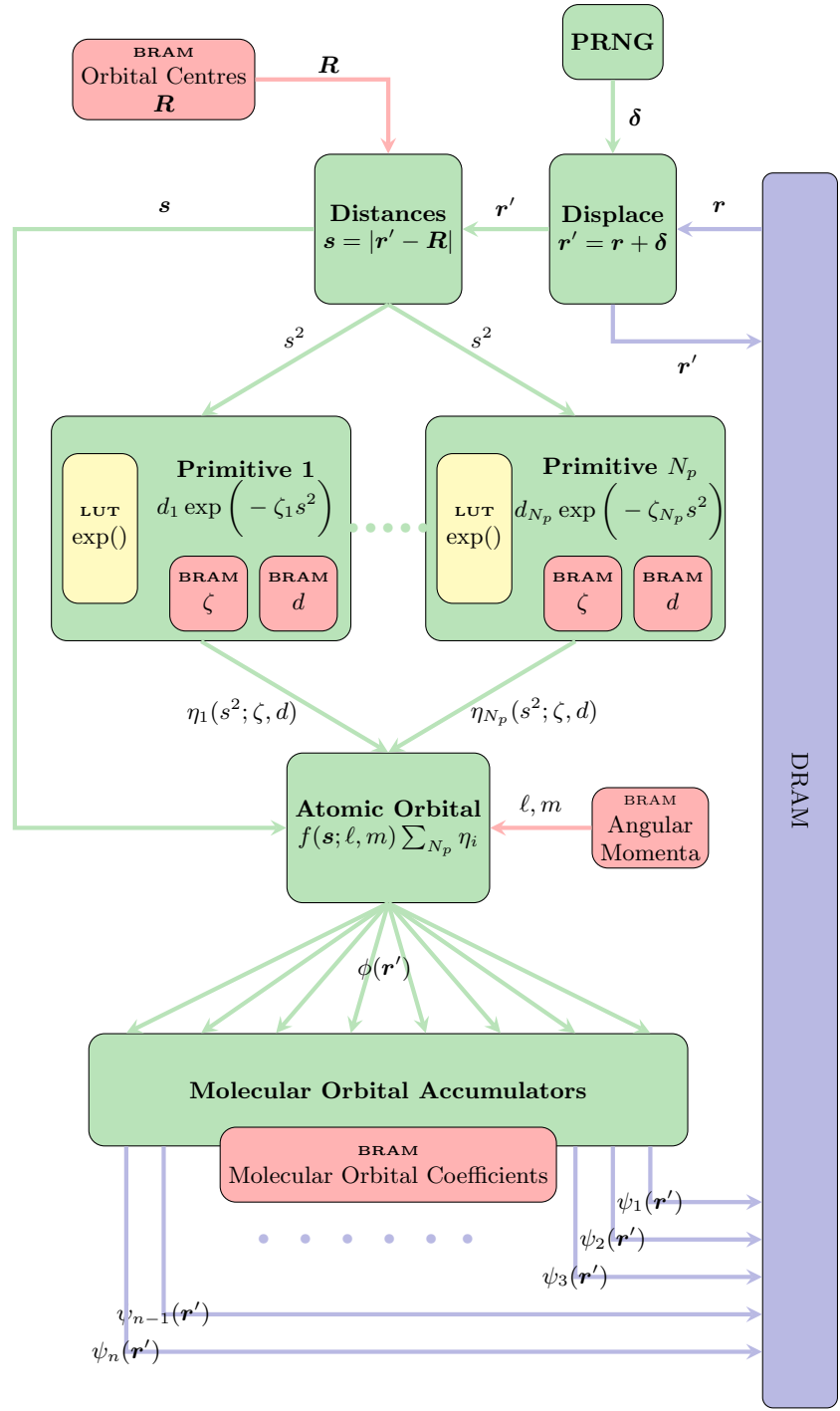
\begin{figure*}[t]
	\tikzsetnextfilename{wavefunction_evaluate}
	\input{wavefunction_evaluate.tikz}
	\caption{\raggedright Wavefunction evaluation dataflow implementation. Red nodes correspond to on-chip memories, yellow nodes to lookup tables. Green nodes are computational kernels, and arrows denote dataflow between elements of the implementation. The blue node represents an on-board addressable DRAM block.}
	\label{fig::WavefunctionEvaluate}
\end{figure*}
In evaluating the trial wavefunction, a number of data parallel regions are amenable to either hardware unrolling or temporal parallelism through the use of stream variables. In constructing a general dataflow solution, our starting point is the identification of a loop which is bounded by some system-independent parameter. The only such loop is given by Equation \eqref{eq::Primitives}, the reduction over primitives to calculate the value of an atomic orbital. For the vast majority of basis sets, the maximum number of primitives associated with an atomic orbital is 6. As such, we choose to unroll the loop over primitives in hardware. In doing so, an adder tree is also required to perform the reduction over primitives.\\
From Equation \eqref{eq::LCAO}, a given atomic orbital $\phi_j(\vec{r})$ is utilised by all molecular orbital LCAOs. Specifically, a fused multiply-add operation is required to accumulate the atomic orbital multiplied by contraction coefficient onto each molecular orbital accumulation. Since the atomic orbital value is available at the end of the pipeline stage which performs the reduction over primitives, an optimal strategy involves immediate use of the atomic orbital value, permitting the calculation of the next atomic orbital in sequence. We consequently unroll the molecular orbital accumulators in hardware. We undertake a similar accumulation over second derivatives of the molecular orbitals since there is the scope for significant data reuse. It should be recognised that unrolling over molecular orbitals and their second derivatives is equivalent to the accumulation of a row of the Slater matrix and laplacian.\\
Our choice for temporal unrolling is now enforced: since we accumulate a single row of the Slater matrix and laplacian at a time, a single electron position vector must be presented to the implementation for $N_{\textrm{AO}}$ consecutive kernel ticks. After this period has passed, the row of the Slater matrix and laplacian will have fully accumulated. These quantities can be offloaded to external memory, the accumulators reset and the next electron streamed through, i.e. the next row of the Slater matrix and laplacian. Since all walkers, and all electrons thereof, must pass through the kernel in this way, we find that our kernel will be required to run for $N_w\times n\times N_{\textrm{AO}}$ ticks. Such counter logic can be realised through three nested counters, over walkers, electrons and atomic orbitals.\\

A number of quantities are fixed over the course of of the calculation: atomic position vectors, molecular orbital coefficients and primitive parameters. Fast access to these quantities is possible through memory-mapping them to the on-chip ROM of the device from the host. However, we only have access to a counter over all atomic orbitals arising in the LCAO for random access, which could result in significant duplication of data should we construct our memory-mapped ROMs in one-to-one correspondence with the atomic orbital counter. In practice, we utilise a set of decoder ROMs, mapping the atomic orbital counter to the address of quantities corresponding to that atomic orbital in the memory-mapped ROMs. The size of the decoder elements need only accommodate the number of bits to required to represent the total number of atomic orbitals, resulting in significant memory savings.\\

The overall schematic for our implementation is given in Figure \ref{fig::WavefunctionEvaluate}. An electron is streamed from DRAM and subjected to a random displacement. As the displaced electron position vector is required by the host for computing the change in potential energy contribution to the local energy, it is streamed back to device DRAM for offload to the host. The vector and square distance between the displaced electron and an atomic orbital centre is computed, with these quantities being used to compute the values of the primitives associated with the atomic orbital under consideration. A reduction amongst the primitives yields the value of the atomic orbital. The atomic orbital value is subsequently passed onto the molecular orbital accumulators, a fused multiply-add being used to accumulate to the LCAO. Finally, upon complete accumulation, the molecular orbital values are passed to device DRAM.\\

Our implementation is reliant upon two key kernels which must also be instantiated within our trial wavefunction kernel: a pseudorandom number generator is required for the displacement of electrons, and an exponential function is required for the calculation of primitive values. Our implementation of these two kernels is discussed in detail in Appendices \ref{app::PRNG} and \ref{app::Exp}.

\subsection{Numerical Precision}

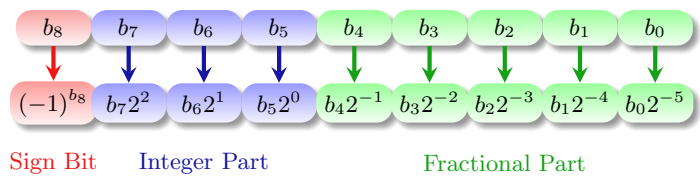
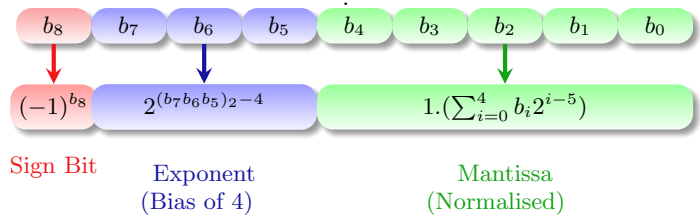
\begin{figure*}[t]
	\begin{subfigure}[t]{\textwidth}
		\tikzsetnextfilename{fixed_point}
		\input{fixed_point.tikz}
		\caption{\raggedright 9-bit fixed point representation. A single ``Sign Bit'' is reserved to represent the signedness of the number. The ``Integer Part'' and ``Fractional Part'' are separated by an implicit radix point. The decimal representation of is given by $(-1)^{b_8}\times (b_7b_6b_5.b_4b_3b_2b_1b_0)_2$}.
		\label{fig::fixedpoint}
	\end{subfigure}
	\begin{subfigure}[t]{\textwidth}
		\tikzsetnextfilename{floating_point}
		\input{floating_point.tikz}
		\caption{}
		\label{fig::floatingpoint}
	\end{subfigure}
	\caption{\raggedright Fixed and Floating Point numerical representations using 9 bits (for the sake of demonstration, i.e. without the implication that 9-bit representations are utilised in our work). Note that these forms are inadequate for intermediary stages in arithmetic operations, where guard digits are used to prevent overflow of the representation.}
\end{figure*}

Floating point arithmetic (see Figure \ref{fig::floatingpoint}) involves several sequentially dependent operations. Addition of floating point numbers, for instance, requires that: the arguments are aligned to the same exponent; the mantissas are summed; the result is normalised (mantissa is shifted and exponent adjusted); and the mantissa is finally rounded to fit within the parameterised bit width. Naturally, these operations can be pipelined, floating point arithmetic thereby being suitable for the out-of-order, deeply pipelined architectures or modern computers. Furthermore, the integration of dedicated ICs, such as floating points units (FPUs), into (or sometimes external to) a CPU permits the offloading of these comparatively cumbersome operations to specialised co-processing units without stalling the CPU.\\
The usage of floating point numerics within an FPGA configuration presents two potential issues: ease of implementation and impact on performance. With regards to the former, a historical obstacle to incorporating floating point numerics into FPGA designs has been the requirement that implementations are written with HDL.  maxj permits the definition of a floating point data type through a single invocation of the method \texttt{dfeFloat()}, taking both the integer and mantissa widths of the floating point representation as arguments. The data type can subsequently be used much like any other data type, with the implementation details handled by the compile-time synthesis.\\

Concerning the second issue, the impact on performance of the FPGA design as a result of floating point arithmetic, the associated latency becomes more problematic than for the CPU. Data streams which are to be combined with the result of a particular floating point operation must be buffered on-chip (as must the actual floating point manipulation stages), so as to accommodate the floating point latency. As a result, logical resources are exhausted without contributing to throughput. Some vendors facilitate the usage of floating point arithmetic through the inclusion of digital signal processors (DSPs) within the fabric of the FPGA. However, such resources are limited, and do not remedy the issues associated with buffering other data streams. Indeed, the usage of DSPs places additional constraints on the place and route, inevitably leading to a potential difficulties in synthesising a particular configuration.\\
One particularly appealing solution is to transition from floating to fixed point arithmetic, the latter being equivalent to integer arithmetic (see Figure \ref{fig::fixedpoint}). One is consequently faced with a number of choices: the width of the representation; whether the numbers are signed or unsigned (i.e. whether two's-complement arithmetic must be supported); and where the radix point partitions integer from fractional parts, amongst other related decisions. To address each of these choices, we must know both the dynamical range and required precision of the quantities to be manipulated. Such matters will be discussed in the following section.\\

\subsubsection{Fixed Point Support for Co-Processing}
There is no native support for fixed point numerics in ISO C99. Nonetheless, it is fairly straightforward to support both fixed point representations and arithmetic through the use of the signed integer data type. However, writing an application which supports fixed point numerics throughout is complicated by a number of issues, particularly when multiple fixed point representations are required:
\begin{enumerate}
	\item Owing to the lack of support for templated functions and operator overloading by the C99 standard, the host code will be bloated and cumbersome.
	\item It is difficult to verify whether an entire application is amenable to fixed point treatment owing to the high-dimensionality of the design space. Supporting fixed point numerics throughout may consequently be an unproductive use of the developer's time should particular operations require floating point treatment.
	\item Use of fixed point arithmetic relies on the construction of bespoke mathematical routines. One consequently foregoes the efficiency of the C standard library routines which have benefited from years of optimisation.
	\item Should the width of the fixed point data type not be byte-aligned, the effective bandwidth of memory transactions will be suboptimal, leading to poor cache efficiency.
\end{enumerate}
By choosing to use fixed point numerics exclusively in the device code, one is able to circumvent the above obstacles to some degree. One particularly appealing aspect of this choice is the constraint of our fixed point design space to an, admittedly complex, kernel. Furthermore, maxj possesses native support for fixed point numerics, including exception handling for under/overflow of the representation along with the inference of representation width from a dynamical range of a quantity. As such, the developer is excused from dealing with low level code optimisation.\\
However, the question remains of the interface between host and device, i.e. at what point the fixed-to-floating point conversions (and vice versa) take place. The simplest interface  involves the offloading of data from the host to the device DRAM in floating point. Upon streaming of data into the DFE, all quantities are cast to and manipulated with an appropriate fixed point representation. Finally, the quantities are cast back to floating point before streaming back to device DRAM. The host is then able to retrieve the data in a natively supported data type.\\
We find this approach somewhat wasteful. Our kernel fully unrolls the Slater matrix and laplacian streams over the dimensionality of the Slater matrices. Consequently, a large number of casts will be performed concurrently, for both input and output streams. Casting between fixed and floating point representations consumes significant on-chip logic resources. As such, a sizeable portion of on-chip resources will be devoted to casting between numerical representations, an overhead associated with the use of the co-processor.  Naturally, such an implementation is best avoided.\\
A preferable scheme materialises when considering the means by which data is transferred between host and device DRAM. Since the MAIA card resides on the PCIe bridge of a compute node, all data transactions between host and device proceed through PCIe. The PCIe driver is instantiated as an IP core in the fabric of the FPGA, requiring data pass through the FPGA over the course of data transactions. Memory transactions which use PCIe are considered a bottleneck of distributed memory systems. The prudent developer of such applications will consequently minimise memory transactions over the course of computation to obtain an optimal implementation.\\
Since data transfer must take place at some point, and the data must pass through the FPGA, it is reasonable to pass the data streams through interface kernels, casting the data as it arrives from the PCIe. A free parameter in our implementation is the width of the stream we cast in the interface kernels. Since the clock frequency of the kernel is known, along with the bandwidth of the PCIe, it is a simple task to evaluate the width of the data stream which saturates the PCIe transfer, and is therefore optimal. For instance, consider a kernel clocked at 100MHz. The bandwidth of a PCIe 2.0 x16 bridge is 8 GB/s. Then, a stream of width 80 bytes saturates the PCIe transfer, equating to 10 double or 20 single precision numbers.\\
A final consideration is the storage of the fixed point data in the device DRAM should the representation not be byte-aligned. It is difficult to find a generic storage width for arbitrary width fixed point data, since DRAM accesses must be burst-aligned. For example, the MAIA card utilised in this work possesses 6 DDR3 channels, each channel having a width of 64 bits. The DRAM can function with a burst size of 8 or 4, resulting in a burst width of 384 or 192 bytes, respectively. Since the burst width will exceed the PCIe saturated width, data must be buffered in the interface kernel for a number of kernel ticks before offloading the buffer to DRAM. Such a scheme is facilitated by using regular widths for the fixed point storage. As such, we choose a storage width of 64 bits. This choice naturally reduces the effective DRAM latency, but for the sake of ease of implementation, we proceed with this parameterisation.\\

\section{Results}

\subsection{Fixed Point}

Use of a single fixed point representation throughout our kernel is not a feasible option. A number of variables possess vastly different dynamic ranges. While a single fixed point representation could in principle be found to accommodate all variables, precision of the representation will necessarily be leveraged for flexibility. As such, our strategy is to track the dynamic range of each variable over the course of a simulation and classify the number of fundamentally different dynamic ranges that arise. To narrow our design space, we will constrain each representation to the same width. We are consequently able to eliminate degrees of freedom from the device space that are only amenable to systematic optimisation strategies.\\
In ascertaining an adequate fixed point representation, we must consider three independent points:
\begin{enumerate}
	\item Whether two's complement is required.
	\item The dynamic range of the variables, such that the integer width can be determined.
	\item The number of fractional bits required to yield a stable calculation.
\end{enumerate} 
The first two points are easily dealt with through analysis. As a further constraint, we concern ourselves specifically with ensembles which have undergone a period of equilibration, thereby further restricting the dynamic range of runtime variables. It is, however, difficult to ascertain the required number of fractional bits without resorting to a full systematic exploration of fractional widths. Our task is complicated all the more given that in establishing the adequacy of a fractional width, a fully converged VMC calculation is required. Naturally such studies are not feasible when using a device simulator, meaning individual DFEs must be synthesised for each fractional width, which again is prohibitively expensive.\\
As such, we employ a scheme requiring a single Monte Carlo step with the device simulator. The displacement of electrons is deactivated, and the kernel computes the Slater matrices and associated derivatives for the same Monte Carlo samples residing on the host. Upon passing these quantities back to the host, we monitor the ensemble averaged local energy and acceptance rate for these dummy moves. If the computed averaged local energy differs from the original value by less than $5\times 10^{-4}$a.u., we deem the implemented fractional bit width adequate. Table \ref{tbl::Precision} shows such results for a varying number of walkers. It is clear that a total fixed point width of 38 bits is the appropriate fixed point representation for our calculation.\\
{\scriptsize{
\begin{table*}[!htb]
\begin{ruledtabular}
\begin{tabular}{c|c|c|c|c|c|c|c|}
	& \multicolumn{7}{c|}{Total Fixed Width}\\
	\cline{2-8}
	$N_w$ & 30 & 32 & 34 & 36 & 38 & 40 & Double \\	
	\hline
	\hline
	128 & -65.1254 & -67.4323 & -68.4111 & -69.6823 & \textbf{-69.6893} & -69.6895 & -69.6895 \\
	256 & -64.2432 & -68.1332 & -68.0236 & -69.7899 & \textbf{-69.7900} & -69.7901 & -69.7901 \\
	512 & -54.9473 & -63.5421 & -64.9021 & -69.7523 & \textbf{-69.7892} & -69.7894 & -69.7894 \\
	1024 & -63.5443 & -61.5423 & -66.3427 & -69.6912 & \textbf{-69.7899} & -69.7899 & -69.7899\\
	2048 & -67.2313 & -68.1998 & -69.7712 & -69.7652 & \textbf{-69.7893} & -69.7895 & -69.7895\\
\end{tabular}
\end{ruledtabular}
\caption{\raggedright Ensemble-averaged local energy over multiple numbers of walkers for varying fixed width representations. A sign bit and the number of integer bits have been dictated by the dynamic range of variables. We find that a total width of 38 bits is adequate to reproduce the ensemble-averaged local energy computed in double precision (final column) to within a tolerance of $5\times 10^{-4}$ a.u.}
\label{tbl::Precision}
\end{table*}
}}
An interesting optimisation which could be undertaken in the future centres around the work of Ceperley and Dewing\cite{ceperley1999penalty}, where a penalty function is applied to random walks with noisy data. We have no reason to assume that truncating a numerical representation introduces a bias into the resultant calculation, and so it appears that one might be able to utilise a shorter fixed width representation and utilise the reported penalty method to compensate for the resulting imprecision. While the acceptance rate will decrease, one can in principle leverage the inefficiency in the Monte Carlo for potential accelerations deriving from the use of a shorter fixed width representation. We plan to investigate the application of the penalty method to this end in the near future.

\subsection{Performance}

The final resource utilisation of our FPGA implementation is given in Table \ref{tbl::Resources}. While our synthesised design is moderately light on consumption of logic and dedicated arithmetic units, the on-chip memory proves to be a limiting factor to increasing the complexity of our design. Both the LUTs of our exponential units and the trial wavefunction parameters, particularly the molecular orbital expansion coefficients, are particularly culpable for such high utilisation. However, given the form of our trial wavefunction, avoiding such overheads is difficult. It is possible to force the MaxCompiler to not pipeline sections of the application, thereby reducing on-chip memory consumption, but at the cost of increased difficulty in the design meeting timing.

\begin{table}[!htb]
{\centering
\begin{tabular}{|c|c|c|}
	\hline
	\textbf{Component} & \textbf{Number Available} & \textbf{Utilisation} \\	
	\hline
	Logic & 262400 & 51.36\% \\
	$18\times 18$ Multipliers & 3926 & 32.32\% \\
	DSP Blocks & 1963 & 36.37\% \\
	On-Chip Memory (M20K) & 2567 & 98.64\% \\
	\hline
\end{tabular}
}
\caption{\raggedright Resource utilisation of our final synthesised FPGA configuration.}
\label{tbl::Resources}
\end{table}

\begin{figure*}[]
	\includegraphics[scale=0.4]{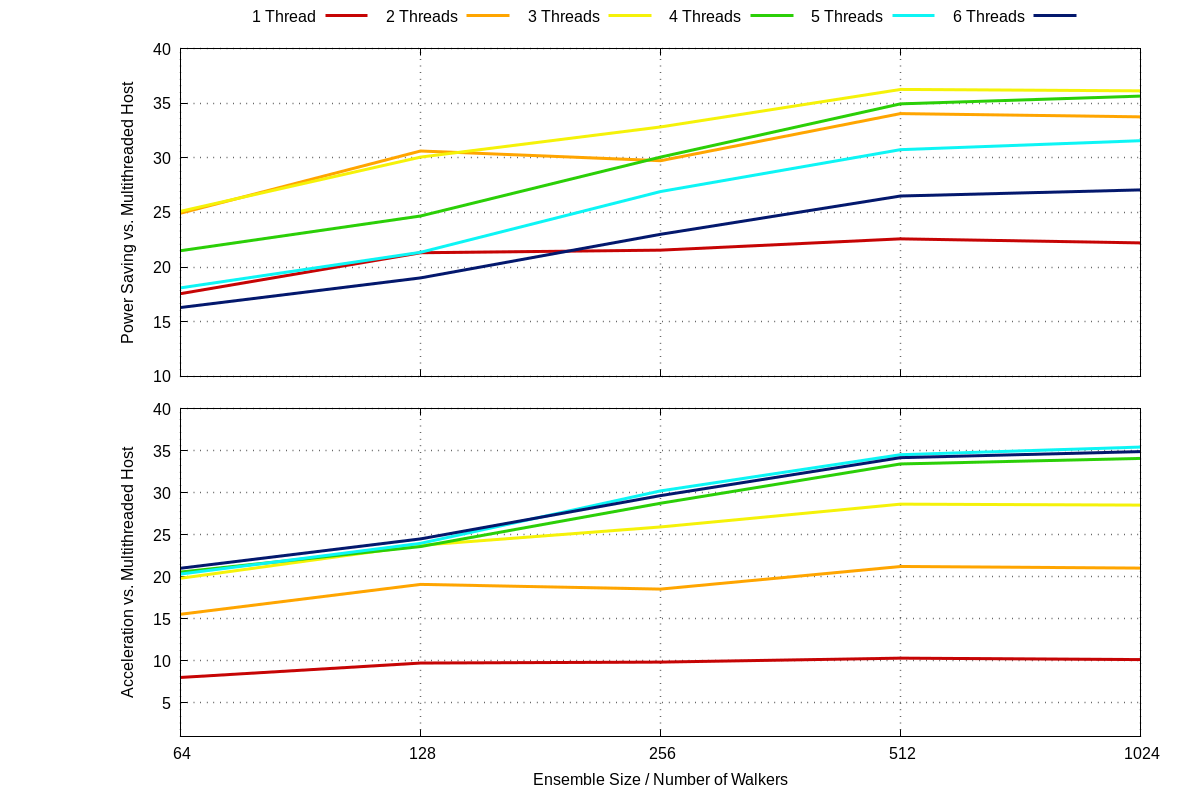}
	\caption{\raggedright Acceleration of the host-device implementation for various ensemble sizes relative to a multithreaded host implementation (lower pane). Performance-adjusted reductions in power consumption are also plotted (upper pane).}
	\label{fig::Performance}
\end{figure*}

Our final results correspond to the overall performance of our application relative to a purely host-bound implementation. We consider a number of ensemble sizes propagated for a fixed number of Monte Carlo steps. In the lower pane of Figure \ref{fig::Performance}, we plot the accelerations relative to the multithreaded host benchmark for a variable number of threads running on the host while the device computes the trial wavefunction.
For small ensemble sizes, we observe a reduction in the performance relative to larger ensemble sizes owing to the overhead associated with using the co-processor. However, for ensemble sizes of 256 walkers and above, we see the improvement in performance converge towards an overall acceleration of roughly $30\times$ relative to the multithreaded host implementation. Across all ensemble sizes, we note that the host need only instantiate 3 or 4 threads of execution to tend towards peak performance.
An interesting additional result is some metric quantifying the comparative power consumptions of the two calculations. While Maxeler Technologies provide a command line utility to query the power consumption of a board, we are not entirely clear on the resolution or accuracy of this quantity. However, in-house testing on a small board with power readings taken at the power outlet reveals the command line utility is in good agreement with the outlet readings. For the MAIA board, we observed a peak power consumption of 27.6W throughout.\\
It is, unfortunately, a little more difficult to establish a power consumption of the host. The operating system is ultimately in control of scheduling, so the overhead associated with other general purpose tasks must be accounted for. A further difficulty is what to include in the power consumption for the host. Typically, only thermal design powers (TDP) are reported by chip manufacturers, which are reported to be in poor correspondence with power consumption at peak operation (peak power consumptions equating to roughly $1.5\times$ that of the reported TDPs.\cite{hennessy2011computer}). The TDP of the Intel Xeon E5-2640 is reported to be 95W. Furthermore, it is unclear whether the power consumption of RAM chips and other peripherals are to be considered.\\
Rather than attempt to speculate on such matters, we compose a metric given the limited data at our disposal:
\begin{equation}
	E = \frac{P_{\textrm{benchmark}} \times S}{P_{\textrm{co-processing}}}
\end{equation}
where $P_{\textrm{benchmark}}$ is the power consumption of the multithreaded host implementation, $P_{\textrm{co-processing}}$ is the power consumption of the device plus that of the number of threads utilised by the co-processing implementation\footnote{When only a subset of physical cores are utilised in the multithreaded co-processing implementation, the power consumption of each core is given by the TDP divided by the number of physical cores.}. $S$ is the speedup of the co-processor relative to the multithreaded host, and $E$ is then the performance-adjusted reduction in power consumption offered by the co-processor. $E$ is plotted in the upper pane of Figure \ref{fig::Performance}. Given that the performance of our application begins to converge to some peak acceleration when the host runs with upwards of three threads, it is unsurprising that the performance-adjusted power consumption is larger for fewer threads than more.\\

\section{Conclusion}

We have ported the computationally expensive trial wavefunction evaluation kernel from a VMC application to an FPGA platform. Through co-processing with a multicore host, we have established that our implementation offers significant benefits in terms of raw compute performance and reduced power consumption. While our VMC is minimal from the perspective of complexity of trial wavefunction, we hope that this work acts to instigate further investigation.\\
Developer time and on-chip resources remain significantly limiting factors in the use of FPGA co-processing for complex scientific applications. However, these limitations are forecasted to become decreasingly problematic with significant effort being directed towards alleviation. The new Intel/Altera Stratix X, for instance, possesses roughly twice the number of on-chip resources as the Stratix V used in this work.\\
Work within our group is currently being directed towards adding further complexity to our implementation, including support for Jastrow factors and localised orbitals. We are also looking to support a DMC application, requiring the first derivative of the trial wavefunction, in addition to support for an FPGA-based Sherman--Morrison inverse updating kernel. We hope that such endeavours will aid in the porting of scientific applications to forecasted exascale platforms.\\

\begin{acknowledgements}
AJWT would like to thank the Royal Society for a University Research Fellowship under grants UF110161 and UF160398 and a Research Grant number RG140728.  This work has been performed as part of the EXTRA\cite{stroobandt2016extra} consortium, supported by the funding from the EU Horizon 2020 research and innovation programme under grant No 671653.  Computations were performed on a Maxeler Galava DFE obtained as part of the Maxeler University Program, as well as the Delorean MAIA cluster as part of the STFC Hartree Centre.  We would also like to thank Dr Jos\'e Gabriel de Figueiredo Coutinho, Dr Timothy Todman and Prof. Wayne Luk of Imperial College London for their advice and use of MAIA compute resources.
\end{acknowledgements}

\section*{Appendices}

\appendix
\section{Pseudorandom Number Generation}\label{app::PRNG}
The Mersenne Twister (MT) is a pseudo-random number generator (PRNG) whose period is given by $2^z - 1$, $z$ being a Mersenne prime\cite{matsumoto1998mersenne}. The most common realisation of the PRNG is parameterised by $z=19937$, hereafter referred to by MT19937. The period of the resultant pseudorandom sequence is therefore adequately long that repetition will not be observed, even over cosmological periods of time with an output frequency of $1 \textrm{fs}^{-1}$.

The recurrence of a sequence is a commonly invoked metric for ascertaining the quality of a random number sequence. The MT19937 is $k$--distributed to 32-bit accuracy, $\forall(1\leq k \leq 623)$, establishing it as a superior PRNG. The pseudorandom sequence output by a MT is given by the recurrence relation
\begin{equation}\label{eq::MTRecurrence}
	\vec{x}_{k+n} = \vec{x}_{k+m} \oplus \big( \vec{x}_k^u || \vec{x}_{k+1}^l \big)\mathcal{A} \,,
\end{equation}
where $\oplus$ and $||$ denote the bitwise $\textrm{XOR}$ and concatenation, respectively, and $n$ is the recurrence of the sequence. We have utilised vector notation to emphasise that the random numbers are each composed of $w$--bits; $\vec{x}^u$ and $\vec{x}^l$ are then the upper and lower bit segments of $\vec{x}$. The length of these segments is ascertained by a single parameter, $r$: $\vec{x}^u$ corresponds to the upper $w-r$ bits of $\vec{x}$, and $\vec{x}^l$ denotes the lower $r$ bits of $\vec{x}$. Finally, $\mathcal{A}$ is a matrix, termed the ``twist-transformation''. The twist-transformation matrix is of rational normal form,
\begin{equation}
	\mathcal{A} = \left[\begin{array}{cc}
    	0 & \mathbf{I} \\
        a_{w-1} & \vec{a}
    \end{array}\right] \,,
\end{equation}
where $\mathbf{I}\in \mathbb{Z}^{w\times w}$ is the identity matrix and $\vec{a}$ is a $w-$bit number, $a_{w-1}$ being the highest-order bit of $\vec{a}$. This form is most convenient for computation, as one is able to evaluate the vector-matrix product of Equation \eqref{eq::MTRecurrence} through simple bitwise operations, without having to explicitly form and store $\mathcal{A}$,
\begin{equation}
	\vec{x}\mathcal{A} = \left\{ \begin{array}{c}
		\vec{x} \gg 1 \\
        (\vec{x} \gg 1)\oplus \vec{a}
	\end{array}\right. \quad \begin{array}{c}
		\textrm{if} \quad x_0 = 0 \\
        \textrm{if} \quad x_0 = 1
	\end{array}
\end{equation}
where $\gg$ is the right-shift operator, and $x_0$ denotes the lowest-bit of $\vec{x}$.

To improve the $k$--distribution of the MT, a ``tempering'' is used for the output pseudorandom numbers, accomplished through application of a tempering matrix, $\mathcal{T}$. As with the twist-transformation matrix, we are spared from explicitly forming/ storing the matrix $\mathcal{T}$ through enforcing that it satisfy a set of pipelined operations:
\begin{align}
	\vec{y}_0 &\leftarrow \vec{x}_{\phantom{0}} \oplus (\vec{x} \gg u) \label{eq::MTTemperingStart} \\
    \vec{y}_1 &\leftarrow \vec{y}_0 \oplus (( \vec{y}_0 \ll s) \hspace{1mm}\&\hspace{1mm} \vec{b}) \\
    \vec{y}_2 &\leftarrow \vec{y}_1 \oplus (( \vec{y}_1 \ll t) \hspace{1mm}\&\hspace{1mm} \vec{c}) \\
    \vec{y}_3 &\leftarrow \vec{y}_2 \oplus (\vec{y}_2 \gg l) \label{eq::MTTemperingEnd}
\end{align}
where $u,s,t,l$ are tempering bit shifts, and $\vec{b},\vec{c}$ are tempering bit masks. The ampersand is used in the conventional sense to denote bitwise $\textrm{AND}$.

The process we have just outlined is most amenable to FPGA implementation. The recurrence relationship of \eqref{eq::MTRecurrence} lends itself to realisation with a ``linear feedback shift register'' (LFSR), where the output of the register containing an initial sequence of $(\vec{x}_0,\vec{x}_1,\hdots,\vec{x}_{k-1}$ is simultaneously fed to the back of the LFSR to construct $\vec{x}_k$, and so on. Furthermore, the pipelined operations outlined in the tempering steps of Equations \eqref{eq::MTTemperingStart} - \eqref{eq::MTTemperingEnd} are suitable for cascaded operations in the fabric of the FPGA. The MT is then a natural candidate as a PRNG for FPGAs, and is our choice for on-chip pseudorandom number generation.

\section{The Exponential Function}\label{app::Exp}
Transcendental functions, by definition, cannot be computed exactly in a finite number of algebraic steps. As such, algorithms which implement transcendental functions must employ some approximation, leveraging speed of convergence for an increase in operational complexity. The exponential function pervades quantum chemistry owing to its utility in simplifying the integrals arising in deterministic methodologies. While useful integral properties are an irrelevance for QMC techniques, linear combinations of exponentials are still commonly used to construct atomic orbitals, as stated in Equation \eqref{eq::Primitives}. As such, in order that our FPGA implementation of the trial wavefunction evaluation routines be as efficient as possible, we have chosen to construct an exponential function which can be queried as rapidly as possible, while consuming minimal on-chip resources. 

Historically, the CORDIC algorithm\cite{andraka1998survey} has been utilised on FPGAs for trigonometric and hyperbolic functions. The identity $\cosh(x) + \sinh(x) = \exp(x)$ can subsequently be used to approximate the exponential function. Since CORDIC is composed entirely of adds and shifts, it leads to a significant reduction in complexity relative to polynomial approximations which require explicit multiplication. However, iterative nature of the CORDIC algorithm renders it suboptimal for our purposes -- the data dependency between iterations is unsuitable for the dataflow implementation we seek to create.

We have instead utilised a lookup table (LUT) approach\cite{doss2004fpga,jamro2007fpga}. While such implementations are memory-intensive, modern FPGA chips have significant amounts of on-chip RAM (of the order of megabytes), and so the cost of storing a few thousand floating point numbers is fairly inconsequential. In order to utilise a LUT-based approach to approximating the exponential function, one must manipulate the identity
\begin{equation}
	\exp( x ) = \Big( 2^{\log_2(e)} \Big)^x = 2^{ x \log_2(e) } \,.
\end{equation}
We proceed to define $x\log_2(e) = y_i + y_f$, where $y_i$ and $y_f$ are the integer and fractional parts, respectively, of the original argument scaled by $\log_2(e)$. We consequently arrive at the expression
\begin{equation}
	\exp(x) = 2^{(y_i + y_f)} = 2^{y_i} \times 2^{y_f} \,.
\end{equation}
A similar partitioning can be undertaken to split $y_f$ down into smaller pieces. Performing this additional partitioning of the fractional part, we can split the approximation down into a series of independent lookups and multiplication of each returned value.

\bibliography{vmc_fpga.bib}

\end{document}

%% file: unpipelined2.tikz
\begin{tikzpicture}

    \def\colorPallete{{"1.00 0 0", "1.0 0.5 0", "1.00 1.00 0.00", "0.8 0.3 0.5", "0.4 0.75 0.00", "0.0 0.75 0.4", "0.0 0.9 0.9", "0.00 0.00 0.75", "0.2 0.0 0.4", "0.0 0.0 0.0" }} ;
    \def\moduleList{{"Input", "$f$", "$g$", "$h$", "Output"}} ;
    
   	\tikzstyle{stagenode} = [draw=none, shade, 
         top color=blue!40, bottom color=blue!5, 
         rounded corners=6pt, minimum width=1cm,
         blur shadow={shadow blur steps=5}, align=center]
   	\tikzstyle{registernode} = [draw=none, shade, 
         top color=red!40, bottom color=red!5, 
         rounded corners=6pt,
         blur shadow={shadow blur steps=5}, align=center]

	\tikzstyle{nd} =
    	[draw=none, shade, 
         minimum width=1.3cm, blur shadow={shadow blur steps=5}]
	\tikzstyle{bl} =
		[draw=none, minimum width=1.5cm, blur shadow={shadow blur steps=5}]

	\node (module_one) [stagenode] at (-0.2, 0) {\textbf{Module} $f$\\ \small{\textit{Latency}: 10}} ;

	\node (module_two) [stagenode, below of=module_one,xshift=1cm,yshift=-0.75cm] {\textbf{Module} $g$\\ \small{\textit{Latency}: 20}} ;

	\node (module_three) [stagenode, below of=module_one,xshift=-1.1cm,yshift=-2.1cm] {\textbf{Module} $h$\\ \small{\textit{Latency}: 10}} ;
    
	\draw[->,ultra thick,-stealth] ([xshift=0.5cm]module_one.south) -- ([xshift=-0.5cm]module_two.north);
	\draw[->,ultra thick,-stealth] ([xshift=-0.5cm]module_one.south) -- ([xshift=0.6cm]module_three.north);
    \draw[->,ultra thick,-stealth] (module_two.south) |- (module_three.east) ;
	\draw[->,ultra thick,-stealth] (-1.75,1.75) node [auto,yshift=0.6cm]{\textbf{Input}}  -- ([xshift=-0.4cm]module_three.north) ;
	\draw[->,ultra thick,-stealth] (module_three.south) -- (-1.3,-5) node [auto,yshift=-0.3cm]{\textbf{Output}} ;
	\draw [->,ultra thick,-stealth] (-1.75,1.75) --  ++(0,-0.4) node [auto, swap]{} -| (module_one.north) ;

	\node (box01) [bl] at (4.8, 0.0) {};
	\node (box02) [bl] at (6.6, 0.0) {};
	\node (box03) [bl] at (8.4, 0.0) {};
	\node (box04) [bl] at (10.2, 0.0) {};

	\node (box12) [bl] at (6.6, -0.5) {};
	\node (box13) [bl] at (8.4, -0.5) {};
	\node (box14) [bl] at (10.2, -0.5) {};
    
	\node (box23) [bl] at (8.4, -1.0) {};
	\node (box24) [bl] at (10.2, -1.0) {};

	\node (box33) [bl] at (8.4, -1.5) {};
	\node (box34) [bl] at (10.2, -1.5) {};

	\node (box44) [bl] at (10.2, -2.0) {};

	\node (box64) [bl] at (10.2, -3.0) {};

	\node (box90) [bl] at (3.0, -4.5) {};
	\node (box91) [bl] at (4.8, -4.5) {};
	\node (box92) [bl] at (6.6, -4.5) {};
	\node (box94) [bl] at (10.2, -4.5) {};

	\node (box10) [bl] at (3.0, -0.5) {};
	\node (box30) [bl] at (3.0, -1.5) {};
	\node (box70) [bl] at (3.0,-3.5) {};
	\node (box84) [bl] at (10.2, -4.0) {};

	\node (box20) [bl] at (3.0, -1.0) {};
	\node (box40) [bl] at (3.0, -2.0) {};
	\node (box60) [bl] at (3.0,-3.0) {};
	\node (box80) [bl] at (3.0, -4.0) {};

	\node (box21) [bl] at (4.8, -1.0) {};
	\node (box31) [bl] at (4.8, -1.5) {};
	\node (box41) [bl] at (4.8,-2.0) {};
	\node (box51) [bl] at (4.8, -2.5) {};
	\node (box71) [bl] at (4.8,-3.5) {};
	\node (box81) [bl] at (4.8, -4.0) {};

	\node (box42) [bl] at (6.6, -2.0) {};
	\node (box52) [bl] at (6.6, -2.5) {};
	\node (box62) [bl] at (6.6,-3.0) {};
	\node (box53) [bl] at (8.4, -2.5) {};
	\node (box63) [bl] at (8.4,-3.0) {};
	\node (box73) [bl] at (8.4, -3.5) {};
	\node (box74) [bl] at (10.2, -3.5) {};
	\node (box83) [bl] at (8.4, -4.0) {};

	\node (box00) [nd, top color=red!40, bottom color=red!5] at (3.0,  0.0) {$x_0$};
	\node (box11) [nd, top color=red!40, bottom color=red!5] at (4.8, -0.5) {$x_0$};
	\node (box21) [nd, top color=red!40, bottom color=red!5] at (6.6, -1.0) {$x_0$};
	\node (box32) [nd, top color=red!40, bottom color=red!5] at (6.6, -1.5) {$x_0$};
	\node (box42) [nd, top color=red!40, bottom color=red!5] at (8.4, -2.0) {$x_0$};
	\node (box53) [nd, top color=red!40, bottom color=red!5] at (10.2, -2.5) {$x_0$};

	\node (box50) [nd, top color=blue!40, bottom color=blue!5] at (3.0, -2.5) {$x_1$};
	\node (box61) [nd, top color=blue!40, bottom color=blue!5] at (4.8, -3.0) {$x_1$};
	\node (box71) [nd, top color=blue!40, bottom color=blue!5] at (6.6, -3.5) {$x_1$};
	\node (box82) [nd, top color=blue!40, bottom color=blue!5] at (6.6, -4.0) {$x_1$};
	\node (box93) [nd, top color=blue!40, bottom color=blue!5] at (8.4, -4.5) {$x_1$};

  	\foreach \module in {0,...,4} {
  		\foreach \time in {0,...,9} {        
        
			\pgfmathsetmacro{\ypos}{-(\time / 2)}
			\pgfmathsetmacro{\xpos}{3+\module * 1.8}
             
    	   	\ifnum\time=0
				\node (label\module) [yshift=0.5cm] at (\xpos,\ypos) {\pgfmathparse{\moduleList[\module]}\pgfmathresult};
			\fi

			\ifnum\module=0
				\node (label\time) [xshift=-1cm,yshift=-0.1cm] at (\xpos,\ypos) {\small$t_\time$};
			\fi
		}
	}

	\draw[<->,ultra thick] ([yshift=-0.5cm]box90.west) -- node [auto,midway,yshift=-0.5cm]{$\mathcal{L} =40$}([yshift=-0.5cm]box94.east) ;
	
    \node(Title) [above of=label0,xshift=1cm,yshift=1.3cm] {\textbf{Unpipelined}} ;
    \node(Latency) [below of=Title,xshift=1.9cm,yshift=0.55cm] {Total Latency, $\mathcal{L} = \ell_f + \ell_g + \ell_h = 40$} ;
    \node(Throughput) [below of=Latency,xshift=-0.35cm,yshift=0.55cm] {Throughput, $\mathcal{T} = \mathcal{L}^{-1} = 0.025^{-1}$} ;

\end{tikzpicture}

%% file: pipelined.tikz
\begin{tikzpicture}

    \def\colorPallete{{"1.00 0 0", "1.0 0.5 0", "1.00 1.00 0.00", "0.8 0.3 0.5", "0.4 0.75 0.00", "0.0 0.75 0.4", "0.0 0.9 0.9", "0.00 0.00 0.75", "0.2 0.0 0.4", "0.0 0.0 0.0" }} ;
    \def\moduleList{{"Input", "$f$", "$g$", "$h$", "Output"}} ;
    
   	\tikzstyle{stagenode} = [draw=none, shade, 
         top color=blue!40, bottom color=blue!5, 
         rounded corners=6pt, minimum width=1cm,
         blur shadow={shadow blur steps=5}, align=center]
   	\tikzstyle{registernode} = [draw=none, shade, 
         top color=red!40, bottom color=red!5, 
         rounded corners=6pt,
         blur shadow={shadow blur steps=5}, align=center]

	\tikzstyle{nd} =
    	[draw=none, shade, 
         minimum width=1.3cm, blur shadow={shadow blur steps=5}]
	\tikzstyle{bl} =
		[draw=none, minimum width=1.5cm, blur shadow={shadow blur steps=5}]

	\node (module_one) [stagenode] at (-0.2, 0) {\textbf{Module} $f$\\ \small{\textit{Latency}: 10}} ;

	\node (module_two) [stagenode, below of=module_one,xshift=1cm,yshift=-0.75cm] {\textbf{Module} $g$\\ \small{\textit{Latency}: 20}} ;

	\node (module_three) [stagenode, below of=module_one,xshift=-1.1cm,yshift=-2.1cm] {\textbf{Module} $h$\\ \small{\textit{Latency}: 10}} ;
    
	\draw[->,ultra thick,-stealth] ([xshift=0.5cm]module_one.south) -- ([xshift=-0.5cm]module_two.north);
	\draw[->,ultra thick,-stealth] ([xshift=-0.5cm]module_one.south) -- ([xshift=0.6cm]module_three.north);
    \draw[->,ultra thick,-stealth] (module_two.south) |- (module_three.east) ;
	\draw[->,ultra thick,-stealth] (-1.75,1.75) node [auto,yshift=0.6cm]{\textbf{Input}}  -- ([xshift=-0.4cm]module_three.north) ;
	\draw[->,ultra thick,-stealth] (module_three.south) -- (-1.3,-5) node [auto,yshift=-0.3cm]{\textbf{Output}} ;
	\draw [->,ultra thick,-stealth] (-1.75,1.75) --  ++(0,-0.4) node [auto, swap]{} -| (module_one.north) ;

	\node (register_one_a) [registernode, below of=module_one,yshift=0.4cm,xshift=-0.5cm,minimum width=0.75cm] {\small{10}} ;
	\node (register_one_b) [registernode, below of=module_one,yshift=0.4cm,xshift=0.5cm,minimum width=0.75cm] {\small{10}} ;

	\node (register_two) [registernode, below of=module_three,yshift=0.4cm, minimum width=0.75cm] {\small{10}} ;
	\node (register_three) [registernode, left of=module_one, minimum width=0.75cm,xshift=-0.5cm] {\small{20}} ;
	\node (register_four) [registernode, below of=register_three, minimum width=0.75cm, yshift=-0.7cm] {\small{20}} ;
	\node (register_five) [registernode, right of=register_four, minimum width=0.75cm] {\small{20}} ;

	\node (box01) [bl] at (4.8, 0.0) {};
	\node (box02) [bl] at (6.6, 0.0) {};
	\node (box03) [bl] at (8.4, 0.0) {};
	\node (box04) [bl] at (10.2, 0.0) {};

	\node (box12) [bl] at (6.6, -0.5) {};
	\node (box13) [bl] at (8.4, -0.5) {};
	\node (box14) [bl] at (10.2, -0.5) {};
    
	\node (box22) [bl] at (6.6, -1.0) {};
	\node (box23) [bl] at (8.4, -1.0) {};
	\node (box24) [bl] at (10.2, -1.0) {};

	\node (box33) [bl] at (8.4, -1.5) {};
	\node (box34) [bl] at (10.2, -1.5) {};

	\node (box43) [bl] at (8.4, -2.0) {};
	\node (box44) [bl] at (10.2, -2.0) {};

	\node (box54) [bl] at (10.2, -2.5) {};
	\node (box64) [bl] at (10.2, -3.0) {};

	\node (box90) [bl] at (3.0, -4.5) {};

	\node (box10) [bl] at (3.0, -0.5) {};
	\node (box30) [bl] at (3.0, -1.5) {};
	\node (box50) [bl] at (3.0, -2.5) {};
	\node (box70) [bl] at (3.0,-3.5) {};
	\node (box84) [bl] at (10.2, -4.0) {};

	\node (box00) [nd, top color=red!40, bottom color=red!5] at (3.0,  0.0) {$x_0$};
	\node (box11) [nd, top color=red!40, bottom color=red!5] at (4.8, -0.5) {$x_0$};
	\node (box21) [nd, top color=red!40, bottom color=red!5] at (4.8, -1.0) {$x_0$};
	\node (box32) [nd, top color=red!40, bottom color=red!5] at (6.6, -1.5) {$x_0$};
	\node (box42) [nd, top color=red!40, bottom color=red!5] at (6.6, -2.0) {$x_0$};
	\node (box53) [nd, top color=red!40, bottom color=red!5] at (8.4, -2.5) {$x_0$};
	\node (box63) [nd, top color=red!40, bottom color=red!5] at (8.4, -3.0) {$x_0$};
	\node (box74) [nd, top color=red!40, bottom color=red!5] at (10.2, -3.5) {$x_0$};

	\node (box20) [nd, top color=blue!40, bottom color=blue!5] at (3.0, -1.0) {$x_1$};
	\node (box31) [nd, top color=blue!40, bottom color=blue!5] at (4.8, -1.5) {$x_1$};
	\node (box41) [nd, top color=blue!40, bottom color=blue!5] at (4.8, -2.0) {$x_1$};
	\node (box52) [nd, top color=blue!40, bottom color=blue!5] at (6.6, -2.5) {$x_1$};
	\node (box62) [nd, top color=blue!40, bottom color=blue!5] at (6.6, -3.0) {$x_1$};
	\node (box73) [nd, top color=blue!40, bottom color=blue!5] at (8.4, -3.5) {$x_1$};
	\node (box83) [nd, top color=blue!40, bottom color=blue!5] at (8.4, -4.0) {$x_1$};
	\node (box94) [nd, top color=blue!40, bottom color=blue!5] at (10.2, -4.5) {$x_1$};

	\node (box40) [nd, top color=green!40, bottom color=green!5] at (3.0, -2.0) {$x_2$};
	\node (box51) [nd, top color=green!40, bottom color=green!5] at (4.8, -2.5) {$x_2$};
	\node (box61) [nd, top color=green!40, bottom color=green!5] at (4.8, -3.0) {$x_2$};
	\node (box72) [nd, top color=green!40, bottom color=green!5] at (6.6, -3.5) {$x_2$};
	\node (box82) [nd, top color=green!40, bottom color=green!5] at (6.6, -4.0) {$x_2$};
	\node (box93) [nd, top color=green!40, bottom color=green!5] at (8.4, -4.5) {$x_2$};

	\node (box60) [nd, top color=yellow!40, bottom color=yellow!5] at (3.0, -3.0) {$x_3$};
	\node (box71) [nd, top color=yellow!40, bottom color=yellow!5] at (4.8, -3.5) {$x_3$};
	\node (box81) [nd, top color=yellow!40, bottom color=yellow!5] at (4.8, -4.0) {$x_3$};
	\node (box92) [nd, top color=yellow!40, bottom color=yellow!5] at (6.6, -4.5) {$x_3$};

	\node (box80) [nd, top color=magenta!40, bottom color=magenta!5] at (3.0, -4.0) {$x_4$};
	\node (box91) [nd, top color=magenta!40, bottom color=magenta!5] at (4.8, -4.5) {$x_4$};
 
  	\foreach \module in {0,...,4} {
  		\foreach \time in {0,...,9} {        
        
			\pgfmathsetmacro{\ypos}{-(\time / 2)}
			\pgfmathsetmacro{\xpos}{3+\module * 1.8}
             
    	   	\ifnum\time=0
				\node (label\module) [yshift=0.5cm] at (\xpos,\ypos) {\pgfmathparse{\moduleList[\module]}\pgfmathresult};
			\fi

			\ifnum\module=0
				\node (label\time) [xshift=-1cm,yshift=-0.1cm] at (\xpos,\ypos) {\small$t_\time$};
			\fi
		}
	}
	
	\draw[<->,ultra thick] ([yshift=-0.5cm]box90.west) -- node [auto,midway,yshift=-0.5cm]{$\mathcal{L} = 60$}([yshift=-0.5cm]box94.east) ;

    \node(Title) [above of=label0,xshift=0.7cm,yshift=1.3cm] {\textbf{Pipelined}} ;
    \node(Latency) [below of=Title,xshift=2.6cm,yshift=0.55cm] {Total Latency, $\mathcal{L} = 3\times\max(\ell_f,\ell_g,\ell_h) = 60$} ;
    \node(Throughput) [below of=Latency,xshift=-0.55cm,yshift=0.55cm] {Throughput, $\mathcal{T} = 3\times\mathcal{L}^{-1} = 0.05$} ;

\end{tikzpicture}

%% file: host_device.tikz
\begin{tikzpicture}

	\tikzstyle{hostnode} = [rectangle, minimum width=3cm, minimum height=3cm, text centered, align=center, rounded corners=6pt, fill=white, draw=black]
	\tikzstyle{cpunode} = [rectangle, minimum width=1cm, minimum height=1cm, text centered, align=center, rounded corners=6pt, fill=myblue!30, draw=black]
	\tikzstyle{memorynode} = [rectangle, minimum width=2.5cm, minimum height=0.7cm, text centered, align=center,
rounded corners=6pt, fill=myyellow!30, draw=black]
	\tikzstyle{fpganode} = [rectangle, minimum width=2.5cm, minimum height=1.3cm, text centered, align=center, rounded corners=6pt, fill=myred!30, draw=black]

	\node (host) [hostnode, label={[yshift=0.2cm]\textbf{Host}}] at (-2, 0) {} ; 
	\node (cpu1) [cpunode] at (-2.8,-0.7) {\small{CPU 1}} ;
	\node (cpu2) [cpunode] at (-1.2,-0.7) {\small{CPU 2}} ;
	\node (ram) [memorynode] at (-2, 0.8) {\small{Memory}} ;        

	\node (device) [hostnode, label={[yshift=0.2cm]\textbf{Device}}] at (2, 0) {} ;        
	\node (dram) [memorynode] at (2, -1) {\small{DRAM}} ;        
	\node (fpga) [fpganode] at (2,0.6) {\small{FPGA}} ;        
        
	\draw[stealth-stealth,double,thick] (cpu1.north) -- ([xshift=-0.8cm]ram.south) ;        
	\draw[stealth-stealth,double,thick] (cpu2.north) -- ([xshift=0.8cm]ram.south) ;        
	\draw[stealth-stealth,double,thick] (dram.north) -- (fpga.south) ;        
	\draw[stealth-stealth,double,thick] ([yshift=0.175cm]fpga.west) -- node[midway,color=black,yshift=0.3cm] {PCIe} (ram.east) ;        
        
\end{tikzpicture}

%% file: non_blocking.tikz
\newsavebox{\bgmaster}
\tikzset{
	background/.style = {
    	rectangle split,
        rectangle split horizontal,
        rectangle split parts = 2,
        rectangle split part fill = {white, black!7.5},
        inner sep = 55pt
    }
}
\tikzset{
	multiplexer/.style = {
    	draw, line width = 0.01mm,
    	trapezium,
    	shape border uses incircle, 
    	shape border rotate = 180
	}  
}
\tikzset{
	arrow/.style = {
    	draw, line width = 0.1mm, scale = 0.1, -stealth
	}  
}
\tikzset{
	hostnode/.style = {
		draw=none, shade, font=\bfseries\sffamily,
        top color=myyellow!40, bottom color=myyellow!5,
        rounded corners=2pt, minimum width=5cm, inner sep=10pt,
        blur shadow={shadow blur steps=5}
    }
}
\tikzset{
	dfenode/.style = {
		draw=none, shade, font=\bfseries\sffamily,
        top color=mygreen!40, bottom color=mygreen!5,
        rounded corners=2pt, minimum width=5cm, inner sep=10pt,
        blur shadow={shadow blur steps=5}
    }
}
\tikzset{
	countnode/.style = {
		diamond, draw=none, shade,
        top color=black!10, bottom color=cyan!5,
        blur shadow={shadow blur steps=5}
    }
}

\pgfdeclarelayer{back}
\pgfsetlayers{back,main}

\begin{tikzpicture}[]

    \node [multiplexer] (mult_1) at (2,-2) {} ;
    \node [multiplexer,right of=mult_1] (mult_2) {} ;
    
    \draw node[left of=mult_1,yshift=0.8cm,xshift=-0.3cm,scale=0.6] (ens_a) {\color{myred}{\texttt{Ensemble\_A}}} ;
    \draw node[below of=ens_a,yshift=0.65cm,scale=0.6] (ens_b) {\color{myblue}{\texttt{Ensemble\_B}}} ;
    \draw node[countnode,below of=ens_b,yshift=0.2cm,xshift=-0.5cm,scale=0.5] (timer) {$t\ \textrm{mod}\ 2$} ;
    
    \draw[arrow,draw=myred,thick] (ens_a.east) -| ([xshift=1.3cm]mult_1.north) ;
    \draw[arrow,draw=myred,thick] (ens_a.east) -| ([xshift=-1.3cm]mult_2.north) ;

    \draw[arrow,draw=myblue,thick] (ens_b.east) -| ([xshift=-1.3cm]mult_1.north) ;
    \draw[arrow,draw=myblue,thick] (ens_b.east) -| ([xshift=1.3cm]mult_2.north) ;

	\draw[arrow,thick] (timer.east) -- node[midway,color=black]{} ++(9,0.0) node [auto, swap]{} |- ([xshift=0.1cm]mult_1.west) ;
	\draw[arrow,thick] (timer.east) -- node[midway,color=black]{} ++(19,0.0) node [auto, swap]{} |- ([xshift=0.1cm]mult_2.west) ;

	\draw node[dfenode,right of=mult_2,xshift=1.5cm,scale=0.5,yshift=-0.7cm] (to_dram) {Offload to DRAM} ;
	\draw node[dfenode,below of=to_dram,scale=0.5,yshift=-0.5cm] (dfe_compute) {Trial Wavefunction Calculation} ;
	\draw node[dfenode,below of=dfe_compute,scale=0.5,yshift=-0.5cm] (from_dram) {Unload from DRAM} ;

	\draw node[hostnode,below of=mult_1,scale=0.5] (metropolis) {Metropolis Accept/Reject} ;
	\draw node[hostnode,below of=metropolis,scale=0.5,yshift=0.2cm] (average) {Ensemble Averages} ;
	\draw node[hostnode,left of=from_dram,below of=average, xshift=1.0cm, yshift=0.05cm,scale=0.5, text centered, align=center] (sync) {Synchronise Host and Device} ;

    \draw[arrow,draw=mygreen,thick] (mult_2.south) |- ++(0,-2cm) node [auto,swap]{} -- (to_dram.west) ;
    \draw[arrow,draw=mygreen,thick] (to_dram.south) -- node[draw=none,midway,scale=0.4,xshift=1cm,align=center] {Stream from\\ DRAM} (dfe_compute.north) ;
    \draw[arrow,draw=mygreen,thick] (dfe_compute.south) -- node[midway,draw=none,scale=0.4,xshift=1cm,align=center] {Stream to\\ DRAM}(from_dram.north) ;
    \draw[arrow,draw=mygreen,thick] (from_dram.west) -- (sync.east) ;
    
    \draw[arrow,draw=myyellow,thick] (mult_1.south) -- (metropolis.north) ;
    \draw[arrow,draw=myyellow,thick] (metropolis.south) -- (average.north) ;
	\draw[arrow,draw=myyellow,thick] (average.south) -- (sync.north) ;

	\draw[arrow,thick] (sync.south) -- node[midway,color=black]{} ++(0,-2.0) node [auto, swap]{} -| node[midway,scale=0.5,yshift=2.5cm,xshift=-0.5cm] {\rotatebox{90}{Increment Counter, $t$}} (timer.south) ;

	\draw node[right of=ens_a,xshift=2cm,yshift=0.5cm] (split_top) {} ;
	\draw node[below of=split_top,yshift=-4cm] (split_bottom) {} ;

	\draw[dashed,thick] (split_top) -- (split_bottom) ;


\end{tikzpicture}

%% file: wavefunction_evaluate.tikz
\begin{tikzpicture}

	\tikzstyle{computenode} = [rectangle, minimum width=1cm, minimum height=2cm, text centered, align=center, rounded corners=6pt,fill=mygreen!30, draw=black]
	\tikzstyle{accumulatenode} = [rectangle, minimum width=8cm, minimum height=1.5cm, text centered, align=center, fill=mygreen!30, draw=black, rounded corners=6pt]
	\tikzstyle{bramnode} = [rectangle, minimum width=1cm, minimum height=1cm, text centered, align=center, fill=red!30, rounded corners=6pt, draw=black]
	\tikzstyle{lutnode} = [rectangle, minimum width=1cm, minimum height=1cm, text centered, align=center, fill=yellow!30, rounded corners=6pt, draw=black]
	\tikzstyle{dramnode} = [rectangle, rounded corners, minimum height=16.5cm, minimum width=1cm, text centered, fill=myblue!30,yshift=-3.5cm, draw=black]

		\node (primitive_1) [computenode, yshift=-2cm, xshift=2cm, minimum height=3cm, minimum width=4cm, label={[xshift=0.6cm,yshift=0.4cm,align=center]center:{\textbf{Primitive 1}\\\small$d_1\exp\bigg(-\zeta_1 s^2\bigg)$}}] at (0, 0) {} ; 
        
        \node (primitive_1_bram_zeta) [bramnode,below of=primitive_1,xshift=0.1cm,yshift=0.13cm] {\tiny{\textbf{BRAM}} \\ $\zeta$} ;
        \node (primitive_1_bram_coeff) [bramnode,right of=primitive_1_bram_zeta, xshift=0.2cm] {\tiny{\textbf{BRAM}} \\ $\vphantom{\zeta}d$} ;
        \node (primtive_1_lut_exp) [lutnode,left of=primitive_1, xshift=-0.35cm, minimum height=2cm] {\tiny{\textbf{LUT}} \\ $\exp()$ } ;

		\node (primitive_n) [computenode, yshift=-2cm, xshift=2cm, minimum height=3cm, minimum width=4cm, label={[xshift=0.6cm,yshift=0.4cm,align=center]center:{\textbf{Primitive $N_p$}\\\small$d_{N_p}\exp\bigg(-\zeta_{N_p} s^2\bigg)$}}] at (5cm, 0) {} ; 
    
    \node[draw=none, right of=primitive_1,color=mygreen!30,xshift=1.5cm] {$\bullet\bullet\bullet\bullet\bullet$} ;

        \node (primitive_n_bram_zeta) [bramnode,below of=primitive_n,xshift=0.1cm,yshift=0.13cm] {\tiny{\textbf{BRAM}} \\ $\zeta$} ;
        \node (primitive_n_bram_coeff) [bramnode,right of=primitive_n_bram_zeta, xshift=0.2cm] {\tiny{\textbf{BRAM}} \\ $\vphantom{\zeta}d$} ;
        \node (primtive_n_lut_exp) [lutnode,left of=primitive_n, xshift=-0.35cm, minimum height=2cm] {\tiny{\textbf{LUT}} \\ $\exp()$ } ;

	\node (compute_distances) [computenode, above of=primitive_1, xshift=2.5cm,yshift=3cm] {\textbf{Distances}\\$\vec{s} = |\vec{r}^\prime - \vec{R}|$} ;
	
	\node (displace) [computenode, right of=compute_distances, xshift=2cm] {\textbf{Displace}\\$\vec{r}^\prime = \vec{r} + \boldsymbol{\delta}$} ;

	\node (random) [rectangle, minimum width=1cm, minimum height=1cm, text centered, align=center, rounded corners=6pt,fill=mygreen!30, draw=black, above of=displace, yshift=1.5cm] {\textbf{PRNG}} ;
    \draw[->,ultra thick,-stealth,color=mygreen!30] (random.south) -- node[midway,color=black,xshift=0.3cm] {$\boldsymbol{\delta}$} (displace.north) ;

	\node (atomic_centre_bram) [bramnode,above of=compute_distances,xshift=-3cm,yshift=1cm] {\tiny{\textbf{BRAM}} \\ Orbital Centres \\ {$\vec{R}$}} ;

	\draw[->,ultra thick,-stealth,color=red!30,] (atomic_centre_bram.east) -| node[midway,color=black,yshift=0.2cm, xshift=-0.8cm] {$\vec{R}$}(compute_distances.north) ;

	\node (dram) [dramnode, right of=primitive_n,xshift=2cm] {\rotatebox{270}{DRAM}} ;
    \draw[->,ultra thick,-stealth,color=myblue!30] ([yshift=7.5cm]dram.west) -- node[midway,color=black,yshift=0.3cm] {$\vec{r}$}(displace.east) ;
    \draw[->,ultra thick,-stealth,color=mygreen!30] (displace.west) -- node[midway,color=black,yshift=0.3cm] {$\vec{r}^\prime$}(compute_distances.east) ;
	\draw[->,ultra thick,-stealth,color=myblue!30] (displace.south) |- node[midway,color=black,yshift=-0.3cm,xshift=1cm] {$\vec{r}^\prime$}([yshift=6cm]dram.west) ;

	\node (atomic_orbital) [computenode, below of=primitive_1, xshift=2.5cm,yshift=-3cm] {\small\textbf{Atomic Orbital}\\$f(\vec{s};\ell,m)\sum_{N_p}\eta_i$} ;
    
    \node (decoders) [bramnode,right of=atomic_orbital,xshift=2.5cm] {\tiny{BRAM} \\ Angular\\Momenta} ;

	\draw[->,ultra thick,-stealth,color=red!30] (decoders.west) -- node[midway,color=black,yshift=0.3cm]{$\ell,m$} (atomic_orbital.east) ;

	\draw[->,ultra thick,-stealth,color=mygreen!30] (compute_distances.west) -- node[midway,color=black,yshift=0.3cm]{$\vec{s}$} ++(-4,0.0) node [auto, swap]{} |- (atomic_orbital.west) ;

    	\draw[->,ultra thick,-stealth,color=mygreen!30] (compute_distances.south) -- node[midway,color=black,yshift=0.3cm] {$s^2$} (primitive_1.north) ;
    	\draw[->,ultra thick,-stealth,color=mygreen!30] (compute_distances.south) -- node[midway,color=black,yshift=0.3cm] {$s^2$} (primitive_n.north) ;

	\draw[->,ultra thick,-stealth,color=mygreen!30] (primitive_1.south) -- node[midway,color=black,xshift=-0.7cm,yshift=-0.2cm] {$\eta_1(s^2;\zeta,d)$} (atomic_orbital.north) ;
	\draw[->,ultra thick,-stealth,color=mygreen!30] (primitive_n.south) -- node[midway,color=black,xshift=0.7cm,yshift=-0.2cm] {$\eta_{N_p}(s^2;\zeta,d)$} (atomic_orbital.north) ;

	\node (molecular_orbitals) [accumulatenode,below of=atomic_orbital, yshift=-2.5cm] {\textbf{Molecular Orbital Accumulators}} ;
    
    \draw[->,ultra thick,-stealth,color=mygreen!30] (atomic_orbital.south) -- ([xshift=-3.5cm]molecular_orbitals.north) ;
    \draw[->,ultra thick,-stealth,color=mygreen!30] (atomic_orbital.south) -- ([xshift=-2.5cm]molecular_orbitals.north) ;
    \draw[->,ultra thick,-stealth,color=mygreen!30] (atomic_orbital.south) -- ([xshift=-1.5cm]molecular_orbitals.north) ;
    \draw[->,ultra thick,-stealth,color=mygreen!30] (atomic_orbital.south) -- ([xshift=-0.5cm]molecular_orbitals.north) ;
    \draw[->,ultra thick,-stealth,color=mygreen!30] (atomic_orbital.south) -- ([xshift=0.5cm]molecular_orbitals.north) ;
    \draw[->,ultra thick,-stealth,color=mygreen!30] (atomic_orbital.south) -- ([xshift=1.5cm]molecular_orbitals.north) ;
    \draw[->,ultra thick,-stealth,color=mygreen!30] (atomic_orbital.south) -- ([xshift=2.5cm]molecular_orbitals.north) ;
    \draw[->,ultra thick,-stealth,color=mygreen!30] (atomic_orbital.south) -- node[midway,color=black,xshift=-1.8cm] {$\phi(\vec{r}^\prime)$} ([xshift=3.5cm]molecular_orbitals.north) ;

    \node (mocoeff_bram) [bramnode,below of=molecular_orbitals] {\tiny{\textbf{BRAM}} \\ Molecular Orbital Coefficients } ;
    
    \draw[->,ultra thick,-stealth,color=myblue!30] ([xshift=3.5cm]molecular_orbitals.south) |- node[midway,color=black] {$\psi_1(\vec{r}^\prime)$} ([yshift=-5.5cm]dram.west) ;
    \draw[->,ultra thick,-stealth,color=myblue!30] ([xshift=3cm]molecular_orbitals.south) |- node[midway,color=black] {$\psi_2(\vec{r}^\prime)$}([yshift=-6cm]dram.west) ;
    \draw[->,ultra thick,-stealth,color=myblue!30] ([xshift=2.5cm]molecular_orbitals.south) |- node[midway,color=black] {$\psi_3(\vec{r}^\prime)$}([yshift=-6.5cm]dram.west) ;
    \node[draw=none, below of=mocoeff_bram,color=myblue!30,xshift=-0.3cm] {$\bullet\quad\bullet\quad\bullet\quad\bullet\quad\bullet\quad\bullet$} ;
    \draw[->,ultra thick,-stealth,color=myblue!30] ([xshift=-3cm]molecular_orbitals.south) |- node[midway,color=black] {$\psi_{n-1}(\vec{r}^\prime)$} ([yshift=-7cm]dram.west) ;
	\draw[->,ultra thick,-stealth,color=myblue!30] ([xshift=-3.5cm]molecular_orbitals.south) |- node[midway,color=black] {$\psi_{n}(\vec{r}^\prime)$} ([yshift=-7.5cm]dram.west) ;

\end{tikzpicture}

%% file: fixed_point.tikz
\begin{tikzpicture}[arrow/.style = {ultra thick,-stealth}]

	\foreach \bit in {0,...,8} {

		\pgfmathtruncatemacro{\result}{8-\bit}
        
        \ifnum\bit=0
			\node (bit\bit) 
        		[draw=none, shade, 
             	 top color=myred!40, bottom color=myred!5, 
                 rounded corners=6pt, minimum width=1cm,
                 blur shadow={shadow blur steps=5}
            	] at (\bit, 0) {$b_\result$};
			\node (value\bit) 
        		[draw=none, shade, 
             	 top color=red!40, bottom color=red!5, 
                 rounded corners=6pt, minimum width=1cm,
                 blur shadow={shadow blur steps=5},
            	 below of=bit0] {$(-1)^{b_8}$};
            \draw[arrow,color=myred] (bit\bit) -- (value\bit) ;

		\else
			\ifnum\bit<4

				\pgfmathtruncatemacro{\power}{3-\bit}
				\node (bit\bit) 
        			[draw=none, shade, 
             	 	 top color=blue!40, bottom color=blue!5, 
                 	 rounded corners=6pt, minimum width=1cm,
                 	 blur shadow={shadow blur steps=5}
            		] at (\bit, 0) {$b_\result$};
                 \node (value\bit) 
        			[draw=none, shade, 
             	 	 top color=blue!40, bottom color=blue!5, 
                 	 rounded corners=6pt, minimum width=1cm,
                 	 blur shadow={shadow blur steps=5},
            	 	 below of=bit\bit] {$b_\result 2^\power$};
                \draw[arrow,color=myblue] (bit\bit) -- (value\bit) ;

			\else

				\pgfmathtruncatemacro{\power}{3-\bit}
				\node (bit\bit) 
        			[draw=none, shade, 
             	 	 top color=green!40, bottom color=green!5, 
                 	 rounded corners=6pt, minimum width=1cm,
                 	 blur shadow={shadow blur steps=5}
            		] at (\bit, 0) {$b_\result$};
                \node (value\bit) 
        			[draw=none, shade, 
             	 	 top color=green!40, bottom color=green!5, 
                 	 rounded corners=6pt, minimum width=1cm,
                 	 blur shadow={shadow blur steps=5},
            	 	 below of=bit\bit] {$b_\result 2^{\power}$};
                \draw[arrow,color=mygreen] (bit\bit) -- (value\bit) ;
    		\fi
		\fi

	}

	\node (signbit) [draw=none, below of=value0,color=myred, yshift=0.2cm] {Sign Bit} ;
	\node (integer) [draw=none, below of=value2,color=myblue, yshift=0.2cm] {Integer Part} ;
	\node (fractional) [draw=none, below of=value6,color=mygreen, yshift=0.2cm] {Fractional Part} ;

\end{tikzpicture}

%% file: floating_point.tikz
\begin{tikzpicture}[arrow/.style = {ultra thick,-stealth}]

	\foreach \bit in {0,...,8} {

		\pgfmathtruncatemacro{\result}{8-\bit}
        
        \ifnum\bit=0
			\node (bit\bit) 
        		[draw=none, shade, 
             	 top color=myred!40, bottom color=myred!5, 
                 rounded corners=6pt, minimum width=1cm,
                 blur shadow={shadow blur steps=5}
            	] at (\bit, 0) {$b_\result$};
			\node (value\bit) 
        		[draw=none, shade, yshift=-0.08cm,
             	 top color=red!40, bottom color=red!5, 
                 rounded corners=6pt, minimum width=1cm,
                 blur shadow={shadow blur steps=5},
            	 below of=bit0] {$\vphantom{\sum_{i=0}^4}(-1)^{b_8}$};
            \draw[arrow,color=myred] (bit\bit) -- (value\bit) ;

		\else
			\ifnum\bit<4

				\pgfmathtruncatemacro{\power}{3-\bit}
				\node (bit\bit) 
        			[draw=none, shade, 
             	 	 top color=blue!40, bottom color=blue!5, 
                 	 rounded corners=6pt, minimum width=1cm,
                 	 blur shadow={shadow blur steps=5}
            		] at (\bit, 0) {$b_\result$};

			\else

				\pgfmathtruncatemacro{\power}{3-\bit}
				\node (bit\bit) 
        			[draw=none, shade, 
             	 	 top color=green!40, bottom color=green!5, 
                 	 rounded corners=6pt, minimum width=1cm,
                 	 blur shadow={shadow blur steps=5}
            		] at (\bit, 0) {$b_\result$};
    		\fi
		\fi

	}

    \node (integer_value) 
    	[draw=none, shade, yshift=-0.08cm,
         top color=blue!40, bottom color=blue!5, 
         rounded corners=6pt, minimum width=3cm,
         blur shadow={shadow blur steps=5},
         below of=bit2] {$\vphantom{\sum_{i=0}^4}2^{(b_7b_6b_5)_2 - 4}$};

     \node (fractional_value) 
         [draw=none, shade, yshift=-0.08cm,
          top color=green!40, bottom color=green!5, 
          rounded corners=6pt, minimum width=5cm,
          blur shadow={shadow blur steps=5},
          below of=bit6] {$1.( \sum_{i=0}^4 b_i 2^{i-5})$};

     \draw[arrow,color=myblue] (bit2) -- (integer_value) ;
     \draw[arrow,color=mygreen] (bit6) -- (fractional_value) ;

	\node (signbit) [draw=none, below of=value0,color=myred, yshift=0.2cm] {Sign Bit} ;
	\node (integer) [draw=none, below of=integer_value,color=myblue, yshift=-0.1cm, text width=2cm] {\centering Exponent \\ \hspace{0.15cm}(Bias of 4)} ;
	\node (fractional) [draw=none, below of=fractional_value,color=mygreen, yshift=-0.1cm, text width=3cm] {\centering Mantissa\\ \hspace{0.45cm}(Normalised)} ;

\end{tikzpicture}